\documentclass[useAMS,usenatbib]{mn2e}

\usepackage[psamsfonts]{amssymb}
\usepackage[dvips]{graphicx}
\usepackage{amsmath,alltt}
\usepackage{multirow}
\usepackage{rotating}
\usepackage{lscape}

\title[On the coexistence of BHs and IMBHs in GCs]{On the coexistence of stellar-mass and intermediate-mass black holes in globular clusters}
\author[Leigh, N. W. C., L\"utzgendorf N., Geller A. M., Maccarone T. J., Heinke C., Sesana A.]{Nathan Leigh$^{1,2}$, 
Nora L\"utzgendorf$^{3}$, Aaron M. Geller$^{4,5}$, Thomas J. Maccarone$^{6}$, 
\newauthor
Craig Heinke$^{1}$, Alberto Sesana$^{7}$
\thanks{E-mail: nleigh@ualberta.ca (NWCL), nluetzge@rssd.esa.int (NL), a-geller@northwestern.edu (AG), thomas.maccarone@ttu.edu (TM), 
heinke@ualberta.ca (CH), alberto.sesana@aei.mpg.de (AE)}\\
$^{1}$Department of Physics, University of Alberta, CCIS 4-183, Edmonton, AB T6G 2E1, Canada \\
$^{2}$Department of Astrophysics, American Museum of Natural History, Central Park West and 79th Street, New York, NY 10024 \\
$^{3}$European Space Agency, Space Science Department, Keplerlaan 1,
2200 AG Noordwijk, The Netherlands \\
$^{4}$Center for Interdisciplinary Exploration and Research in Astrophysics (CIERA) \&  Dept. of Physics and Astronomy, \\ Northwestern Uni
versity, 2145 Sheridan Rd, Evanston, IL 60208, USA \\
$^{5}$Department of Astronomy and Astrophysics, University of Chicago, 5640 S. Ellis Avenue, Chicago, IL 60637 \\
$^{6}$Department of Physics, Texas Tech University, Box 41051, Lubbock TX 790409-1051, USA \\
$^{7}$Max Planck Institute for Gravitational Physics, Albert Einstein Institute, Am M\"uhlenberg 1, 14476, Golm, Germany}
\begin{document}

\pagerange{\pageref{firstpage}--\pageref{lastpage}} \pubyear{2011}

\maketitle

\label{firstpage}

\begin{abstract}
In this paper, we address the question:  What is the probability of stellar-mass black hole (BH) binaries 
co-existing in a globular cluster with an intermediate-mass black hole (IMBH)?  Our results suggest that 
the detection of one or more BH binaries can strongly constrain the presence of an IMBH in most Galactic 
globular clusters.  More specifically, the detection of one or more BH binaries could strongly indicate 
against the presence of an IMBH more massive than $\gtrsim 10^3$ M$_{\rm \odot}$ in roughly 80\% of the 
clusters in our sample.  To illustrate this, we use a combination of $N$-body simulations and analytic 
methods to weigh the rate of formation of BH binaries against their ejection and/or disruption rate via 
strong gravitational interactions with the central (most) massive BH.

The eventual fate of a sub-population of stellar-mass BHs (with or without binary companions) is for all 
BHs to be ejected from the cluster by the central IMBH, leaving only the most massive stellar-mass BH 
behind to form a close binary with the IMBH.  During each phase of evolution, we discuss the rate of 
inspiral of the central BH-BH pair as a function of both the properties of the binary and its host cluster.
\end{abstract}


\section{Introduction} \label{intro}


The topic of whether black holes (BHs) in globular clusters (GCs) exist has long been a topic 
of debate.  Arguments have been made in favour of the presence of not only stellar-mass BHs 
($\sim$ 10-10$^2$ M$_{\odot}$), but also intermediate-mass BHs (IMBHs; 
$\sim$ 10$^3$-10$^5$ M$_{\odot}$) in some clusters.  The former are descended directly 
from the evolution of massive ($\gtrsim$ 25 M$_{\odot}$) stars, and may or may not 
experience a kick upon formation due to an asymmetry in the subsequent supernova explosion 
\citep[e.g.][]{fryer12}.  IMBHs, on the other hand, could form from runaway collisions in 
the cluster core \citep{portegieszwart04}, 
the direct collapse of gas in a protocluster, or accretion onto a central stellar-mass 
BH \citep{leigh13b}.

The first theoretical suggestions of some kind of connection between GCs and BHs
were that the GCs might be quasar remnants \citep{wyller70}.  This suggestion has long
since been abandoned, but the discovery of a disproportionate number of X-ray sources in GCs
\citep{clark75} led to the suggestion that these clusters might host black holes of $\sim 1000 M_\odot$
accreting from the intracluster medium \citep{bahcall75}.  At nearly the same time, it
was found that the velocity dispersion rises toward the center of M15, one of the X-ray bright
clusters \citep{newell76}.

%

As quickly as the evidence in favor of IMBHs in GCs came together, opposing explanations to account 
for their signatures were introduced.  The X-ray 
emitting globular clusters all soon showed Type I X-ray bursts \citep{grindlay76}, which were then 
explained as thermonuclear runaways on the surfaces of neutron stars \citep{ayasli82}.  The 
increase in central mass to light ratio in globular clusters was argued to be consistent with mass 
segregation putting the heavy remnants into the very centers of GCs \citep{illingworth77}.  
White dwarfs and neutron stars have very large mass-to-light ratios in the optical, and hence they can mimic 
a central IMBH if enough of them are present.  Similar arguments for 
\citep{gerssen02} and against \citep{baumgardt03} the presence of an intermediate mass black hole 
in M15 appeared almost 30 years later.

More recently, two primary methods of indirect detection have been the focus of attempts to identify 
IMBHs in GCs.  These are the fitting of dynamical models to look for evidence of cusps in the central 
velocity dispersion profiles \citep[e.g.][]{gebhardt05,noyola08,pasquato09,vandermarel10,anderson10,luetzgendorf13a}, 
and X-ray and especially radio continuum 
observations to look for evidence of gas accretion onto a central massive black hole 
\citep[e.g.][]{maccarone08,strader12a,haggard13}.  
%
Both kinematic- and accretion-based approaches are subject to caveats when interpreting 
the observations \citep[e.g.][]{umbreit13}.  For example, although the detection of stellar density and 
kinematic cusps in GCs could be indicative of an IMBH, cusp stars near the very centre 
of the cluster could have moderately anisotropic orbits, and this could also mimic the evidence for 
an IMBH \citep[e.g.][]{ibata09}.  Similarly, an upper limit can be placed on the Eddington ratio of 
an accreting IMBH in a GC.  But, if the gas content is poorly known \citep[e.g.][]{bellazzini08}, a low gas 
accretion rate could be 
indicative of a very low-density interstellar medium (ISM), as opposed to a low IMBH mass 
\citep[e.g.][]{wrobel11}.  
Other methods of detecting IMBHs have also been explored.  For example, \citet{drukier03} proposed that 
individual fast-moving stars could be created in GCs hosting IMBHs, and that such hypervelocity 
stars could be observable using the Hubble Space Telescope.  \citet{gill08} also showed that an 
IMBH can potentially quench mass segregation, and cause the average stellar mass to vary only 
modestly as a function of the clustercentric radius.  Later, \citet{sesana12} also suggested that 
fast-moving millisecond pulsars in the halo of our Galaxy could provide indirect 
evidence for a substantial population of IMBHs in GCs, since such high 
velocities are difficult to reproduce via ``standard'' formation mechanisms.

Currently, the observational evidence in favour of stellar-mass BHs 
existing in GCs is more compelling than for IMBHs.  To detect these BHs, they must be accreting 
from a binary companion.  For example, \citet{maccarone07} first found an accreting BH in a GC 
associated with the giant elliptical galaxy NGC 4472 in the Virgo Cluster.  Although its
precise mass is not known, the x-ray luminosity is sufficiently high that
it cannot be anything other than a BH in such an old stellar population.  This result was expanded 
upon by \citet{zepf07}, who also discussed the possible implications for a central 
IMBH.  Soon thereafter, 
\citet{shih10} reported an accreting BH in a GC hosted by the giant elliptical galaxy NGC 1399 
located at the centre of the Fornax Cluster.  
More recently, \citet{strader12b} reported the detection of two flat-spectrum 
radio sources in the Galactic GC M22.  The authors presented 
compelling evidence that these are accreting stellar-mass BHs, the first of 
their kind to be detected in a Galactic GC.  These two detections arguably imply 
the presence of $\sim 5 - 100$ stellar-mass BHs in M22, and could be 
indicative of stellar-mass BHs existing in other Galactic GCs as well. Indeed, 
\citet{chomiuk13} recently reported a candidate BH x-ray binary in the Galactic 
GC M62.  

The exact numbers and masses of stellar-mass 
BHs in GCs is still being heavily debated in the literature.  Most, if not
all, BHs should be located in the cluster core at the present-day ages of Galactic GCs.
This is because they 
should drift into the core due to mass segregation on the shortest time-scales, if
they are not already born there.  However, when it comes to BHs, what happens in the core
rarely stays in the core.  Strong gravitational interactions between BHs can result
in the dynamical ejection of all but one or two BHs \citep{sigurdsson93b,kulkarni93}.

More specifically, the \citet{spitzer69} instability shows that a heavy stellar component 
will dynamically decouple into a central ``sub-cluster'' if it has a critical combination of 
both total mass and ratio of masses of individual objects in the heavy component to the light 
component.  Since the crossing time, and especially the relaxation time, is shorter for the 
sub-cluster than for a typical globular cluster, this sub-cluster can be expected to evaporate 
itself on much less than a Hubble time, ejecting the BHs dynamically \citep{sigurdsson93b}.  
Furthermore, binary interactions may accelerate the process \citep{portegieszwart00}.  
Given that all the bright (L$_X >$ 10$^{36}$ erg/sec) Galactic globular cluster X-ray sources have shown 
strong evidence for being neutron stars, a lore developed that the Spitzer instability really did 
lead to the evacuation of black holes from GCs.  Only with the discoveries of objects 
with strong evidence for being globular cluster black holes both in other galaxies 
\citep{maccarone07,maccarone11} and in the Milky Way's clusters \citep{strader12b,chomiuk13} 
was this topic carefully revisited.  An important point was raised, namely, that the Spitzer instability 
should never drive the complete evacuation of a cluster's black holes \citep[e.g.][]{moody09}, since 
the criterion for the 
Spitzer instability to work includes a total mass requirement on the heavy component 
\citep{maccarone11}.  

More recent numerical work has shown that the black hole retention 
fraction is similar to the neutron star retention fraction 
\citep[e.g.][]{moody09,morscher13}.  For 
example, \citet{sippel13} performed $N$-body simulations to 
match the absolute and dynamical age of M22.  The authors argue that multiple BHs 
are retained at a cluster age of 12 Gyr provided the parent cluster has an 
extended core radius.  These results were quickly expanded upon by 
\citet{breen13} 
and \citet{heggie13}, who argue that it should take on the order of ten relaxation 
times for the entire BH sub-population to be ejected, and that clusters with 
present-day half-mass relaxation times above $\sim$ 1 Gyr are the ones that should 
retain an appreciable population of BHs at 12 Gyr \citep{downing10}.

In this paper, we consider the co-existence of a population of stellar-mass BHs and a 
central IMBH in a GC.  
In Section~\ref{fate}, we use 
$N$-body simulations to demonstrate that the most massive stellar-mass BH becomes 
bound to the IMBH on a relatively short timescale.  
We then discuss the evolution of the IMBH-BH binary in Section~\ref{evolve}.  Using 
analytic methods, we weigh the probability of actually 
detecting BH X-ray binaries in GCs hosting an IMBH in Section~\ref{BHbinary}.  Finally, 
our main conclusions are summarized in Section~\ref{summary}.

\section{The fate of a population of stellar-mass BHs in a GC hosting an IMBH} \label{fate}

Because of their relatively large masses, populations of stellar-mass BHs in GCs rapidly sink to the cluster 
centre due to mass segregation \citep{vishniac78,larson84,breen13,heggie13}.  Here, 
they can undergo strong gravitational interactions with other BHs, possibly leading to their ejection from 
the cluster and hence a progressive depletion of the BH sub-population \citep{spitzer69,sigurdsson93b,morscher13}.   
If a central IMBH is present, this process is accelerated \citep{luetzgendorf13b}.  In this case, the 
stellar-mass BHs are ejected upon undergoing strong gravitational interactions with the IMBH.  

\begin{figure*}
\begin{center}
\includegraphics[width=\textwidth]{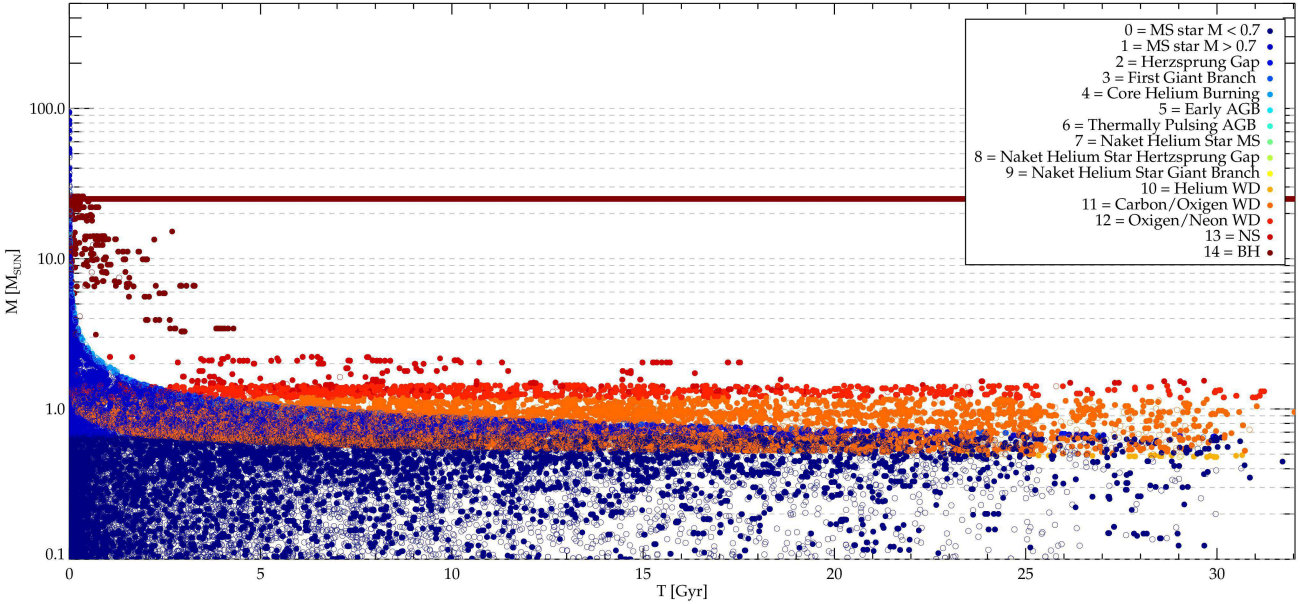} 
\end{center}
\caption[Bound and unbound encounters with the central IMBH over time]{Plot showing 
every bound (solid circles) and unbound (open circles) encounter that occurs with 
the central IMBH over the course of the cluster evolution.  Time is shown in Gyr 
on the x-axis, whereas the mass (in M$_{\odot}$) of the object that interacts 
with the IMBH is shown on the y-axis.  The colour-coding of the circles correspond 
to the different stellar types, and are shown in the inset at the top right.  
The dark red circles correspond to stellar-mass BHs, the last of which (ignoring the 
BH bound to the IMBH in a close binary at the cluster centre) is ejected from the 
cluster at $\sim 4$ Gyr.
\label{fig:fig1}}
\end{figure*}

The above scenario is illustrated in Figure~\ref{fig:fig1}, which depicts the 
results of an $N$-body simulation for cluster evolution, 
%
performed using NBODY6 \citep{aarseth99}.  We use the same initial 
conditions and model setup as in \citet{luetzgendorf13b} with N = 131~072 stars initially.  
Additional simulations are also performed for N $\sim$ 64~000 and N $\sim$ 32~000 stars (not 
shown in Figure~\ref{fig:fig1}), adopting again the initial conditions in \citet{luetzgendorf13b}.
In all simulations, the IMBH mass, binary and remnant retention fractions are M$_{\bullet} =$ 0.01M$_{cl}$, 
f$_{bin} =$ 0.0, f$_{BH} =$ 0.3, f$_{NS} =$ 0.1, f$_{WD} =$ 1.0 \citep{luetzgendorf13b}.  
Importantly, the IMBH mass is chosen to be 1\% of the total cluster mass at 12 Gyr.  For lower IMBH masses, 
the IMBH is typically ejected from the cluster due to strong gravitational interactions with other 
stellar remnants and/or massive stars (if any remain).  Specifically, the IMBH masses are 833 M$_{\odot}$, 
415 M$_{\odot}$ and 207 M$_{\odot}$ for the 128k, 64k and 32k models, respectively.  
This is larger than predicted by extending the M-$\sigma$ relation observed for super-massive black holes in 
galactic nuclei \citep[e.g.][]{ferrarese00}, but more reasonable if the \textit{initial} GC masses were much 
larger than their present-day values \citep{kruijssen13}. 

Note that we do not include any primordial binaries in these simulations to minimize the computational 
cost, but do allow for the dynamical formation of (typically wide) binaries.\footnote{Our $N$-body 
simulations under-estimate the rate of BH binary formation by not including primordial binaries.  This issue is 
addressed by combining the results of our 
$N$-body simulations with analytic estimates for the rate of BH binary formation.}  Here, we are primarily 
concerned with interactions between stellar-mass BHs and the central IMBH.  Importantly,
all BHs are treated as point particles and we do not consider gravitational wave emission in our
simulations.  Thus, the orbital separation of the central IMBH-BH binary only evolves due to
encounters with other objects, and it never undergoes a merger.  We will return to this
issue in the subsequent section.

Since we are interested in direct encounters with the central IMBH we modified the code 
to output every close encounter with the IMBH.  For every time step the distance 
of each object to the IMBH is calculated and, if smaller than the encounter radius $r_{enc} = 10^{-3}$ pc, 
their orbital parameters obtained and saved. Bound objects can be identified by their orbital 
energy or their recurrence in the following time steps. Specifically, an encounter is classified
as bound (unbound) if, at the first timestep after the object has passed its point of closest approach, 
its kinetic energy is less (greater) than the absolute value of its orbital energy (with the IMBH) 
at that distance from the IMBH.  We follow the cluster evolution for $\gtrsim$ 12 Gyr in all our 
simulations.

Figure~\ref{fig:fig1} depicts all encounters that occurred during the cluster lifetime.  Open 
symbols mark unbound encounters and filled symbols bound encounters.  
As is clear from Figure~\ref{fig:fig1}, the most massive stellar-mass BH in the 
cluster becomes closely bound to the central IMBH within the first $\lesssim 100$ Myr 
of cluster evolution.  The remaining black holes then undergo strong encounters with 
this central BH-IMBH binary, until eventually all BHs but the central pair 
are ejected from the cluster.  This characteristic behaviour is seen in all our simulations, 
however there is some stochasticity to the time required for all BHs to be ejected
from the cluster \citep[e.g.][]{downing10,heggie13}, and it can occur that another BH is 
exchanged into the IMBH-BH binary, ejecting the first companion from the cluster in the process.  
The final BH, not bound to the IMBH, is ejected after only 4 Gyr in the simulation shown in 
Figure~\ref{fig:fig1}.  This 
timescale ranges from $\sim$ a few 100 Myr in our simulations with 32k particles to several 
Gyr in our simulations with 128k particles (see Figure~\ref{fig:fig5}).  
In the most massive GCs ($\gtrsim$ 10$^6$ M$_{\odot}$) with the longest relaxation times, however, 
this timescale could be much longer and even exceed a Hubble time \citep[e.g.][]{downing10}.  
Importantly, the heaviest BHs tend to be 
ejected first, since they have the shortest relaxation times.  Hence, they drift close to 
the cluster centre on the shortest timescales, where they undergo a strong interaction 
with the central IMBH-BH pair.  Thus, we 
expect the ejection times for the heaviest BHs to typically be the shortest.  


\section{Evolving the central IMBH-BH binary} \label{evolve}

\begin{figure}
\begin{center}
\includegraphics[width=\columnwidth]{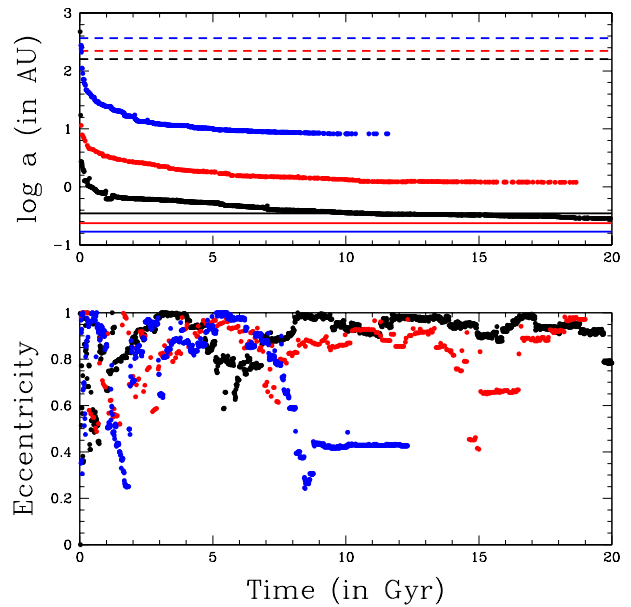}
\end{center}
\caption[Time evolution of the IMBH-BH orbital separation and eccentricity in three $N$-body
simulations]{Plot showing the time evolution (in Gyr) of the orbital
separation (top panel) and eccentricity (bottom panel) of the central IMBH-BH binary.  The
results are shown for three different $N$-body simulations, each beginning with a different
number of particles, namely N $\sim$ 32k (blue), 64k (red) and 128k (black).
The dashed and solid lines correspond to a $=$ a$_{\rm h}$ and a $=$ a$_{\rm GW}$,
respectively, calculated for each of our three $N$-body simulations assuming an eccentricity
of 0.5 for the IMBH-BH binary.
\label{fig:fig2}}
\end{figure}

Next, we turn our attention to the evolution of the orbital properties of the central IMBH-BH binary.  
Using the $N$-body simulations described in Section~\ref{fate}, we begin with a qualitative discussion 
of the evolution of the IMBH-BH binary prior to the final inspiral, at which point GW emission takes over 
as the dominant mechanism for orbital decay.  Here, however, we also show the results for 
simulations that begin with N $\sim$ 32k and N $\sim$ 64k particles initially, in addition to the
case with N $\sim$ 128k shown in Figure~\ref{fig:fig1}.  The time evolution of
the semi-major axes and eccentricities of the IMBH-BH binaries formed in these simulations are
shown in Figure~\ref{fig:fig2}.  The key features to note are the steady decrease in orbital
separation due to hardening encounters with the surrounding stellar population, and that the
orbital eccentricty is close to unity for much of the binary's lifetime (we will come back to
this below).  

Initially, it is dynamical friction from the surrounding population
acting on the individual binary components that drives the reduction in orbital separation
\citep{milosavljevic01}.  Eventually, scattering encounters take over as the dominant hardening
mechanism.  The critical semi-major axis at which this transition occurs is called the hard binary 
separation a$_{\rm h}$, and is approximately given by
the boundary for the IMBH-BH binary to be classified as dynamically hard
in the core \citep{heggie75}, or:
\citep{quinlan96,merritt13}:
\begin{equation}
\label{eqn:hard-soft}
a_{\rm h} = \frac{M_{\rm 2}}{M_{\rm 12}}\frac{r_{\rm m}}{4},
\end{equation}
where M$_{\rm 12} =$ M$_{\rm 1} +$ M$_{\rm 2}$ is the total mass of the IMBH-BH binary, with
M$_{\rm 2} <$ M$_{\rm 1}$.  The distance r$_{\rm m} =$ GM$_{\rm 1}$/$\sigma^2$ is the influence
radius of the more massive IMBH.  Thus, the hard binary separation is inversely proportional to 
the square of the central velocity dispersion $\sigma^2$, which tends to increase with increasing 
cluster mass due to the virial theorem (along with the expected IMBH mass).  This is illustrated in 
Figure~\ref{fig:fig2} via 
the dashed lines, as well as our $N$-body simulations since, at any given time, the semi-major axis 
of the central IMBH-BH binary is larger for smaller cluster masses, and hence IMBH masses.  

From Equation~\ref{eqn:hard-soft}, the semi-mjor axis 
a$_{\rm h}$ of the IMBH-BH binary at which scattering interactions take-over as the dominant 
mechanism of orbital decay could range anywhere from a few AU in the most massive GC cores (with the 
highest velocity dispersions and the most massive IMBHs) to on the order of 100 AU in the least massive GC 
cores.  For example, adopting component masses M$_{\rm 1} =$ 10$^3$ M$_{\rm \odot}$ and 
M$_{\rm 2} =$ 10 M$_{\rm \odot}$ 
and assuming a central velocity dispersion of 10 km s$^{-1}$, Equation~\ref{eqn:hard-soft} gives 
$\sim 22$ AU roughly independent of M$_{\rm 1}$ if M$_{\rm 1} \sim$ M$_{\rm 12} \gg$ M$_{\rm 2}$.

We expect the eccentricity of the IMBH-BH binary to evolve over time, maintaining a relatively high 
eccentricity (ignoring GW emission).  
Indeed, an initially non-circular orbit is the most likely scenario when the IMBH-BH binary first 
forms \citep{milosavljevic01,valtonen06}.  

\begin{figure}
\begin{center}
\includegraphics[width=\columnwidth]{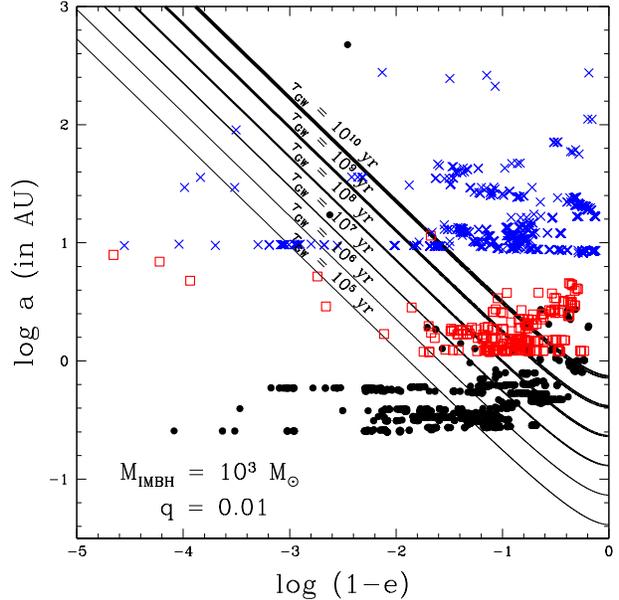}
\end{center}
\caption[Curves of constant $\tau_{\rm GW}$ for an IMBH-BH binary in the semi-major
axis-eccentricity-plane]{Relation between the initial eccentricity and the initial semi-major axis
for an IMBH-BH binary that will merge due to gravitational wave emission within the indicated time
$\tau_{\rm GW}$, calculated using Equation~\ref{eqn:GRemission}.  The IMBH and BH masses are set to
M$_{\rm IMBH} =$ 10$^3$ M$_{\odot}$
and M$_{\rm BH} =$ 10 M$_{\odot}$, respectively, giving a mass ratio q $=$ 0.01.  The different
line widths correspond to different time-scales (10$^5$, 10$^6$, 10$^7$, 10$^8$, 10$^9$ and 10$^{10}$ years) 
for a merger to occur due to GW emission, such that 
the time-scale increases with increasing line width.  The evolution of the IMBH-BH binaries in our $N$-body
simulations are shown in the $\log$(1-e)-$\log$(a)-plane by the different points.  Specifically, the black solid
circles, red open squares and blue crosses correspond to the results from our 128k, 64k and 32k simulations,
respectively.  Note that, in all simulations, the semi-major axis of the IMBH-BH binary decreases monotonically
with time, and the time between points varies but is on the order of 10 Myr.
\label{fig:fig3}}
\end{figure}

Stellar encounters are not the only physical mechanism affecting the evolution of the IMBH-BH binary.  Compact 
binaries lose orbital energy to gravitational waves.  The timescale for a binary to merge from gravitational wave 
emission is:
\begin{equation}
\begin{aligned}
\label{eqn:GRemission}
\tau_{\rm GW} = 3.3 \times 10^8 \frac{(1+q)^2}{q} \Big( \frac{a}{1 AU} \Big)^4 \\
\Big( \frac{M_1+M_2}{10^3 M_{\odot}} \Big)^{-3} (1-e^2)^{7/2}\mbox{ years},
\end{aligned}
\end{equation}
where a and e are the initial semi-major axis and eccentricity of the IMBH-BH binary, and 
q $=$ M$_{\rm 1}$/M$_{\rm 2}$ is the binary mass ratio.  This allows us to introduce the concept of a critical 
semi-major axis a at which a binary of mass ratio q and eccentricity e$_{\rm GW}$ coalesces within 
a specified time.  This is illustrated in Figure~\ref{fig:fig3}, where 
we show curves of constant $\tau_{\rm GW}$ in the semi-major axis-eccentricity-plane.  The evolution of the 
IMBH-BH binaries in our $N$-body simulations are also shown by the different points.  Specifically, the 
solid black circles, open red squares and blue crosses correspond to the results of our 128k, 64k and 
32k simulations, respectively.  In all simulations, the timescale for coalscence due to GW emission drops 
well below 1 Myr, due primarily to the very high eccentricities reached by the IMBH-BH binary (we will come 
back to this below).  

\begin{figure}
\begin{center}
\includegraphics[width=\columnwidth]{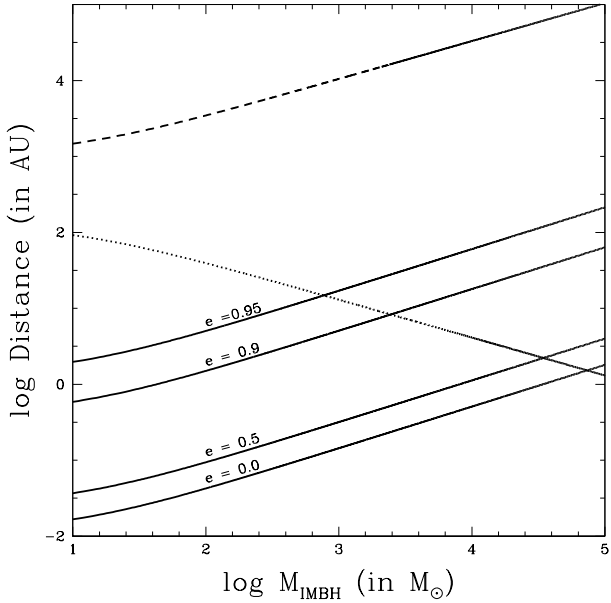}
\end{center}
\caption[Characteristic distances from the central IMBH as a function of IMBH mass]{The 
parameters a$_{\rm h}$ (dotted line) and a$_{\rm GW}$ (solid lines) denote the critical 
semi-major axes at which hardening interactions and gravitational wave emission, respectively, 
take over as the dominant source of orbital decay.  These are shown in AU as a function 
of IMBH mass (in M$_{\odot}$), 
calculated from Equations 19 and 20 in \citet{sesana10} assuming $\gamma = 1.0$ (appropriate for clusters 
with cores, as opposed to cusps).  The companion BH mass is set to 
10 M$_{\odot}$, and we show the results for orbital eccentricities e $=$ 0, 0.5, 0.9, 0.95.  Also 
shown by the dashed line is the influence radius of the central IMBH, as given by Equation 3 
in \citet{sesana10}.
\label{fig:fig4}}
\end{figure}

Eventually, a critical semi-major axis a$_{\rm GW}$ is reached, at which point the timescale for coalescence 
due to GW emission is comparable to the (instantaneous) timescale for coalescence due to scattering 
interactions.  
The dependences of the critical semi-major axes a$_{\rm h}$ and a$_{\rm GW}$ on the mass of the 
central IMBH are shown in Figure~\ref{fig:fig4} by the dotted and solid lines, respectively, 
for different eccentricities and assuming a 10 M$_{\odot}$ BH companion.  For very low eccentricities, 
a$_{\rm GW}$ is roughly two orders of magnitude lower than a$_{\rm h}$ ($\sim 0.1$ and $\sim 10$ AU, 
respectively) for IMBH masses $\sim 10^3$ M$_{\odot}$.  However, for very high eccentricities 
approaching unity, we find a$_{\rm GW} \sim$ a$_{\rm h}$ at the same IMBH mass.  

The key point to take away from Figures~\ref{fig:fig3} and~\ref{fig:fig4} is that, as the eccentricity 
increases, the efficiency of 
GW emission increases, acting to circularize the IMBH-BH binary on potentially shorter timescales than 
scattering interactions can increase it.  Thus, if the eccentricity becomes sufficiently
high, the rate of orbital decay due to GW emission could dominate over scattering interactions, even
at large semi-major axes.  

To summarize, the orbital eccentricity of the central IMBH-BH binary reaches very
high values at relatively small semi-major axes in as little as a few Gyr in all our simulations, due to
scattering interactions with the surrounding stellar population.  This significantly reduces the
timescale for the IMBH-BH binary to merge due to GW emission to $\sim$ a few to a few hundred Myr
(depending on the exact semi-major axis and eccentricity), which is not accounted for in our
$N$-body simulations.  Thus, we expect many IMBH-BH binaries in GC cores to merge
well within a Hubble time, in particular those with large masses and located in dense GCs.  Naively, this 
could reduce the timescale for the depletion of the stellar-mass BH population.  These issues should be 
properly addressed in future $N$-body models that 
incorporate a proper treatment of the General Relativistic evolution of the central IMBH-BH binary.

\section{Can a BH binary co-exist with an IMBH?} \label{BHbinary}


In the subsequent sections, we address in more detail whether or not any BHs should remain in a 
cluster hosting an IMBH at the present-day and, if so, whether or not they might harbour binary companions.  

\subsection{Will any BHs remain at the present-day?} \label{remain}

\begin{figure}
\begin{center}
\includegraphics[width=\columnwidth]{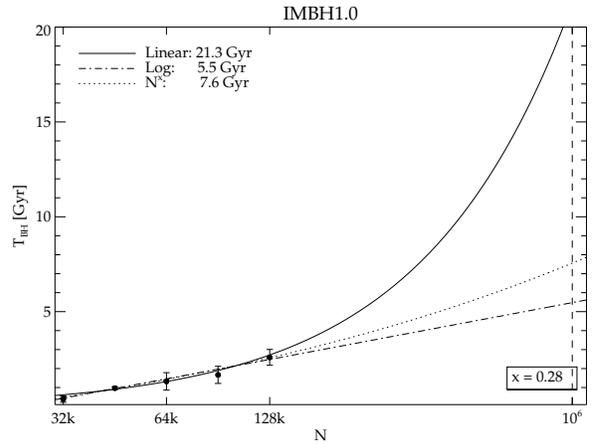}
\end{center}
\caption[The time until all but the last three BHs have been ejected from the cluster as a
  function of the initial total number of stars]{The time until all but the last three BHs
have been ejected from the cluster (T$_{\rm BH}$) is shown as a function of the initial total
number of stars in the cluster ($N$) for all three $N$-body models.  Three
different fits to the data are shown by the solid, dotted and dot-dashed lines, as
described in the inset.  The indicated times in the inset correspond to the time
at which all (but three) BHs have been ejected in a cluster with N $=$ 10$^6$ stars
initially.  Additional $N$-body simulations taken from \citet{luetzgendorf13b} are
also shown.
\label{fig:fig5}}
\end{figure}

In general, for the same initial BH retention fractions, the timescale for all BHs to be ejected increases
with increasing cluster mass, since the relaxation time increases with increasing cluster mass (and the 
number of initial BHs likely increases with cluster mass).  This is 
illustrated in Figure~\ref{fig:fig5}.  Here, T$_{\rm BH}$ indicates the time until all but
the last three stellar-mass BHs have been ejected from the cluster, since there is considerably
more stochasticity between simulations in the time until all but \textit{two} BHs have
been ejected.  Three different fits to the data are shown, however more models will be needed 
to better constrain the fits at higher initial $N$-values.  To first order, the linear-$N$ and 
logarithmic-$N$ fits can be regarded as upper and lower limits, respectively, for T$_{\rm BH}$.  
All three fits show an increase in T$_{\rm BH}$ with increasing $N$.  Assuming
T$_{\rm BH} \propto N^x$, where $N$ is the total number of particles initially and x is a free
parameter, Figure~\ref{fig:fig5} shows that more than three BHs should remain at 12 Gyr in a
cluster with $N =$ 10$^6$, provided x $\lesssim$ 0.6.

If some BHs remain in the cluster when the first stellar-mass BH companion merges with the IMBH due to 
gravitational wave emission, the IMBH will capture another BH companion on a timescale that is 
comparable to the timescale for BH ejection.  As illustrated in Figure~\ref{fig:fig1}, the most massive 
BHs tend to be ejected first.  Hence, each time the IMBH captures an additional BH companion after 
undergoing a merger, the new BH is likely less massive than its predecessor(s).  The lower companion mass 
increases the timescale for orbital decay due to GW emission at a$_{\rm GW}$, which contributes to 
decreasing the rate at which the IMBH merges with other stellar-mass BHs over time.  Apart 
from this, we do not expect mergers of BHs with the central IMBH 
to significantly change the presented picture for the dynamical depletion of a cluster's BH 
sub-population.  

\subsection{Will any primordial BH binaries remain in the cluster at the present-day?} \label{primordialBHs}

In this section, we discuss only BHs born with a binary companion, which we call primordial BH binaries, and 
post-pone the discussion of their dynamical formation to the subsequent section.  

Most, and probably all, primordial BH binaries born in the cluster core should be disrupted 
by the central IMBH by the present-day, or ejected from the cluster.  This is because the timescale 
for two-body relaxation 
in the core is less than a Hubble time in even the most massive Galactic GCs, but is $\ll 1$ Gyr
in a typical GC \citep{harris96}.  The central two-body relaxation time provides a rough guide (we will 
return to this in the next section, where we calculate a more accurate timescale) for
the timescale on which massive objects already in the core sink sufficiently deep within the cluster 
potential to enter 
the sphere of influence of the IMBH-BH binary, where the force of gravity from the IMBH becomes
significant.  For example, using Equation 2.12 in \citet{merritt13} and assuming
a central velocity dispersion of 10 km s$^{-1}$, the influence radius is $\sim$ 0.04 pc for a
10$^3$ M$_{\odot}$ IMBH, or $\sim$ 0.4 pc for a 10$^4$ M$_{\odot}$ IMBH.  
Even if a BH binary is able to survive a direct encounter with an IMBH or, more likely, an IMBH-BH binary 
(in the subsequent section, we quantify this probability), there is usually ample time for subsequent 
interactions to occur, since the core and even half-mass relaxation times are typically much shorter 
than the cluster age.  This significantly increases the probability that any primordial BH binaries will 
be disrupted by the present-day cluster age, or even ejected from the cluster.  While
it is possible that some BH binaries could survive an interaction with the central IMBH-BH binary and
survive to the present-day by being ejected to the cluster outskirts where the timescale for
dynamical friction to operate is very long, such BH binaries should be rare and should most
likely still be located in the cluster outskirts at the present-day.

Can primordial BH binaries born outside of the core survive to the present-day?  To address this 
question, we consider two different types of BHs, motivated by the most recent theoretical work and defined 
according to the mass of their progenitor.  The 
first type (Type I) of BH consists of the lowest mass BHs in the cluster (with a progenitor mass 
$\lesssim 40$ M$_{\odot}$, although the exact limit is highly uncertain), 
and experiences a significant kick upon formation \citep{fryer12}.  We therefore assume
that most BHs of Type I are formed as isolated objects, and do not have a binary companion
at birth.  This is because, even if the BH progenitor had a binary companion when it
underwent a supernova (SN) explosion, the imparted kick is highly likely to dissociate the 
binary.\footnote{The presence of a binary companion at the time of formation may help to retain 
BHs in clusters, since it can occur that a large fraction of the imparted kick energy goes 
into unbinding the binary.  Also, the additional mass of the companion reduces the mass 
segregation time of the object, and natal kicks are less likely to eject BHs if they occur 
deeper within the gravitational potential of the cluster.}
Thus, if BHs of Type I are to have binary companions, they must typically capture them via 
some dynamical mechanism.  
Importantly, we expect most BHs to be of Type I (before considering 
ejection from the cluster due to natal kicks) given that the initial mass function (IMF) observed 
in the field and young star-forming regions rises toward lower masses \citep[e.g.][]{kroupa02}.  
With that said, it is difficult to quantify this difference considering the considerable 
uncertainties in the IMF in massive clusters at the high-mass end, as well as the magnitude 
of their natal kicks.

The second type (Type II) of BH constitutes the most massive stellar remnants in the cluster (with the 
possible exception of an IMBH) with a 
progenitor mass $\gtrsim 40$ M$_{\odot}$, and does not experience a kick upon formation, since the
progenitor collapses directly to a BH without a supernova explosion, and with little
mass-loss \citep{fryer12}.\footnote{We note that, as far as we know, no observational evidence 
exists for BHs descended directly from stellar progenitors with masses $\gtrsim$ 40 M$_{\odot}$.  Moreover, 
we expect significant mass loss due to 
stellar winds in such massive star progenitors near the Eddington limit, and this could also significantly 
lower the final BH mass relative to the initial progenitor mass.  We use this division in mass, given originally 
in \citet{fryer12}, as 
a guide to help explore the possible presence of a sub-population of especially massive stellar-mass BHs that 
do not experience natal kicks.}  Thus,
it is entirely possible, and even likely, that BHs of Type II are born with a binary
companion, given the high binary fractions among massive stars \citep[e.g.][]{sana09,sana11}.  However,
given the large masses of these BHs, they have very short dynamical friction timescales,
even at very large clustercentric radii.  To check this, we calculate
the distance from the cluster centre at which the dynamical friction timescale is equal to
12 Gyr for a 40 M$_{\odot}$ object, for 42 non-core collapsed GCs using data taken from the
ACS Survey for GCs \citep{sarajedini07}.  The GC sample, single-mass King model fits and method for
our procedure are described in detail in \citet{leigh11b}.  We find that the distance from the
cluster centre is $\gtrsim 95$\% of the tidal radius in most GCs.  Even for
a GC as massive as $\omega$ Cen, BHs of Type II must reside well beyond the cluster
half-mass radius at birth in order to have mass segregation times that exceed the age
of the cluster, and avoid drifting into the core within a Hubble time.  Thus, most, if not all, BH binaries 
of Type II should end up in the core well before the present-day cluster age, at which point they 
should undergo a strong gravitational interaction with the central IMBH-BH binary within a 
central relaxation time. 

We conclude that most, if not all, primordial BH binaries should have either been disrupted 
by the central IMBH-BH binary or ejected from the cluster by the present-day.  

\subsection{How can isolated BHs acquire a binary companion?} \label{formation}

Ignoring kicks imparted by dynamical interactions in the core, most BHs will spend at least some of 
their lifetimes in the cluster core.  The exceptions are Type II BHs born at 
very large clustercentric radii (well beyond the half-mass radius and, for all but the most massive
GCs with the longest relaxation times, likely very near to the tidal
radius), and Type I BHs kicked to comparably large clustercentric radii by their natal 
supernova kick or an encounter with the central IMBH.  Any Type I BHs born at large clustercentric 
radii are likely to be ejected from the cluster due to their natal SN kick.  The key point that most BHs 
will potentially spend significant time in the core is important since, as we will explain below, 
dynamical mechanisms 
for BH binary formation require a high stellar density to be efficient.  Thus, \textit{most BH binaries 
should form in the core} (see below for a more quantitative estimate).  We consider only dynamical formation 
channels in the core for BH binaries, and do 
not address their possible formation via binary evolution.  Thus, we ask:  How long for a single BH 
to acquire a binary companion via some dynamical mechanism?  


First, we consider an exchange interaction between an isolated BH and a binary, which occurs in 
the cluster core.  Hence, we 
use the single-binary encounter time given in \citet{leigh11a}, multiplied by the number 
of single stars in the core:
\begin{equation}
\begin{gathered}
\label{eqn:coll1+2}
\tau_{\rm sb} = 1.4 \times 10^{9} \Big( \frac{1}{f_{\rm b}} \Big) \Big(\frac{10^5 pc^{-3}}{n_{\rm 0}} \Big) \\
 \Big(\frac{v_{\rm m}}{5 kms^{-1}} \Big) \Big(\frac{0.5 M_{\odot}}{m_{\rm BH}} \Big) \Big( \frac{1 AU}{a} \Big)\mbox{ years},
\end{gathered}
\end{equation}
where f$_{\rm b}$ is the core binary fraction, n$_{\rm 0}$ is the core number density, v$_{\rm m}$ is 
the root-mean-square stellar velocity in the core, m$_{\rm BH}$ is the mass of the interloping BH and 
a is the average binary semi-major axis in AU.
This gives the time for a \textit{specific} BH to encounter \textit{any} 
binary, as opposed to the time for \textit{any} BH-binary pairing to occur.  The latter 
timescale is shorter than the former by a factor N$_{\rm BH}$, where N$_{\rm BH}$ is the 
number of BHs remaining in the core at the present-day cluster age.  Thus, technically, we should also 
divide the 
single-binary timescale given in \citet{leigh11a} by the number of BHs N$_{\rm BH}$ in order to 
obtain the time for \textit{any} BH to acquire a binary companion via an exchange interaction.  We 
will come back to this below.  

Assuming a core 
radius of 1 pc, a central mass density of 10$^5$ M$_{\odot}$ pc$^{-3}$, a central velocity dispersion of 
10 km s$^{-1}$, an average stellar mass of 0.8 M$_{\odot}$, a binary fraction of 10\% and 
an average binary semi-major axis of 1 AU,\footnote{This is slightly shorter than the semi-major axis 
corresponding to the hard-soft boundary, which is $\sim 7$ AU.} we calculate a time of 0.6 Gyr 
for a \textit{specific} 10 M$_{\odot}$ BH to encounter \textit{any} binary.  This is shorter than 
the core relaxation time calculated above for our example cluster, for which we obtain 1.5 Gyr 
(following \citet{binney87}).  For comparison, if for the same example cluster we adopt a more modest 
binary fraction of 3\%, which 
is observed in the cores of a few massive MW GCs \citep[e.g.][]{milone12}, we calculate an encounter
time of 1.9 Gyr for a 10 M$_{\odot}$ BH to encounter any binary, which is longer than the central relaxation 
time.  We note that the encounter time is even shorter if divided by a factor N$_{\rm BH}$ to 
obtain the time for \textit{any} BH to undergo an encounter.  On the other hand, the time for an \textit{exchange} 
to occur is longer than the encounter time by a factor of $\sim$ a few \citep[e.g.][]{leonard89}, 
since not every encounter results in an exchange event.  Technically, \textit{this} is the relevant 
timescale for comparison to the timescale on which 
most stellar-mass BHs (already in the core) will undergo a strong encounter with the central IMBH.\footnote{More 
accurately, the exchange timescale should be compared to the timescale for diffusion of BHs into the 
central IMBH-BH binary's loss-cone.  The loss-cone can be roughly defined as the collection of orbits with 
a periastron distance that is roughly equal to the IMBH-BH binary's semi-major axis.  However, this 
timescale is comparable to, albeit typically slightly shorter than, the central relaxation time.  We 
will return to this important issue in the subsequent section, where the relevant loss-cone timescale 
is calculated directly.}  
This will either dissociate any binary companions the BHs may have acquired, or eject them from the cluster.  We note that 
binaries formed from exchanges (and tidal capture events; see below) typically are not compact, and only 
very compact binaries are 
expected to survive an encounter with the central IMBH-BH binary \citep[e.g.][]{sigurdsson93a}.  
%

A second possible formation scenario for BH binaries involves an isolated BH acquiring 
a binary companion in the cluster core via tidal capture, most likely of a single main-sequence 
(MS) star.\footnote{A BH 
could also collide directly with a red giant star, forming an ultraluminous X-ray binary in the 
process \citep{ivanova05a}.  We do not consider this mechanism directly, and instead assume that 
all single stars lie on the MS.}  We 
adopt the tidal capture timescale given in \citep{kalogera04}, with an additional factor 
(1 - f$_{\rm b}$)$^{-1}$, where f$_{\rm b}$ is the binary fraction in the cluster core.  This factor 
adjusts for the fact that we are not concerned with interactions 
involving binaries in the tidal capture scenario.\footnote{It is certainly possible, however, 
that tidal capture occurs during interactions involving binaries.  Thus, our derived timescale 
can be regarded as an upper limit.}  This gives:
\begin{equation}
\begin{gathered}
\label{eqn:tc}
\tau_{\rm tc} = 10^9 \Big( \frac{1}{1-f_{\rm b}} \Big) \Big( \frac{10^5 pc^{-3}}{n_{\rm 0}} \Big) \\
 \Big(\frac{v_{\rm m}}{10 kms^{-1}} \Big) \Big(\frac{5 R_{\odot}}{r_{\rm tc}} \Big) 
\Big( \frac{10 M_{\odot}}{m_{\rm BH}} \Big)\mbox{ years},
\end{gathered}
\end{equation}
where r$_{\rm TC}$ is the tidal capture radius.

For the example cluster described in 
the previous paragraph and assuming a core binary fraction f$_{\rm b} =$ 0.1, we calculate 
for a 10 M$_{\odot}$ BH a tidal capture time of 0.3 Gyr (assuming r$_{\rm tc} =$ 5 R$_{\odot}$).  
This is shorter than the 
timescale for a BH to acquire a binary companion via an exchange interaction in the core by a 
factor $\sim 2$.  Thus, in this example cluster, the tidal capture timescale becomes longer 
than the binary exchange timescale for core binary fractions $\gtrsim$ 20\%.  Despite these 
arguments in favour of the tidal capture scenario, it is important 
to note that tidal capture may very well lead to the complete disruption of a MS star 
\citep{kalogera04}.  More detailed modeling of these interactions will be required in 
future studies to fully address this issue.

\subsection{Can the detection of BH binaries constrain the possible presence of an IMBH?}

\begin{figure}
\begin{center}
\includegraphics[width=\columnwidth]{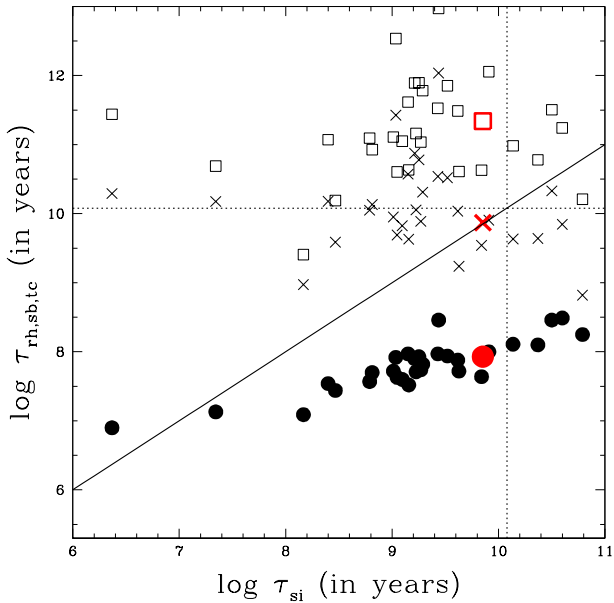}
\end{center}
\caption[The strong interaction timescale versus the half-mass relaxation time, single-binary encounter time and 
tidal capture time for a sample of Galactic GCs]{The half-mass relaxation time $\tau_{\rm rh}$ (filled cirles), single-binary 
encounter time $\tau_{\rm sb}$ (open squares) and tidal capture time $\tau_{\rm tc}$ (crosses) are plotted as a function of 
the strong interaction time $\tau_{\rm si}$.  As described in the text, 
all timescales are calculated for a \textit{single} 10 M$_{\odot}$ BH.  The half-mass relaxation time is multiplied
by the ratio of the average stellar mass to the mass of the BH, and we adopt 0.1 AU
for the average binary semi-major axis in calculating the binary encounter time.  The GC sample and
data are described in Section 2 of \citet{leigh13c}.  The solid line denotes the one-to-one line, and the dotted lines 
show the 12 Gyr mark on both axes.  The red enlarged points correspond to M22 (NGC 6656).
\label{fig:fig6}}
\end{figure}

To evaluate whether or not the detection of one or more BH binaries can be used to argue against the 
simultaneous presence of an IMBH, we first must calculate the time for BHs to be ejected from the cluster 
due to strong gravitational interactions with the central 
IMBH-BH binary.  This depends on the efficiency of diffusion of BHs into the loss-cone of the IMBH-BH 
binary, which we define as the ensemble of orbits with a periastron distance that is on the order 
of the semi-major axis of the IMBH-BH binary.  Objects on loss-cone orbits are expected to undergo 
sufficiently strong encounters with the IMBH-BH binary that they are typically ejected from the cluster 
within a crossing time.  We assume that the binary loss-cone is always full, such 
that objects on intersecting orbits with the IMBH-BH binary are continually re-supplied as they 
are ejected.  This is a reasonable assumption in GCs, since the time for BHs to be ejected (see below) is 
typically comparable to the half-mass relaxation time (see Figure~\ref{fig:fig6}), or:
\begin{equation}
\label{eqn:t-rh}
\tau_{\rm rh} = 1.7 \times 10^5 N^{1/2}\Big( \frac{r_{\rm h}}{1 {\rm pc}} \Big)^{3/2} \Big( \frac{1 M_{\odot}}{\bar{m}} \Big)^{1/2}\mbox{ years},
\end{equation}
where N is the total number of objects in the cluster, r$_{\rm h}$ is the half-mass radius and $\bar{m}$ is the 
average stellar mass.  

In this approximation, the relevant timescale corresponds to that for all BHs, \textit{once in the core}, 
to undergo strong interactions with 
the IMBH-BH binary, which we call the ``strong interaction timescale''.  This should depend 
on the IMBH-BH semi-major axis, which is not accounted for by the central relaxation time.  To 
calculate this timescale, we follow the prescription of \citet{sesana12}, with a few minor adjustments.  
In particular, based on Equation 13 in \citet{sesana12}, we write the strong interaction timescale as:
\begin{equation}
\begin{gathered}
\label{eqn:strongtime}
\tau_{\rm si} = 4.7 \times 10^7 \Big( \frac{r_{\rm h}}{1 pc} \Big)^3 \\
 \Big( \frac{v_{\rm m}}{10 kms^{-1}} \Big)^3 \Big( \frac{\bar{m}}{1 M_{\odot}} \Big)^2 \\
 \Big( \frac{10 M_{\odot}}{m_{\rm BH}} \Big)^3 \Big(  \frac{10^3 M_{\odot}}{m_{\rm IMBH}} \Big)\mbox{ years},
\end{gathered}
\end{equation}
where r$_{\rm h}$ is the half-mass radius in parsecs, v$_{\rm m}$ is the cluster root-mean-square velocity, m$_{\rm IMBH}$ 
denotes the mass of the central IMBH and m$_{\rm BH}$ is the mass of its binary companion (both in M$_{\odot}$).  
We have confirmed that the timescale given by Equation~\ref{eqn:strongtime} agrees quite well with the results 
of our $N$-body simulations.  For example, Equation~\ref{eqn:strongtime} predicts $\tau_{\rm si} + \tau_{\rm rh} \sim$ a 
few Gyr for our $N$-body model with 128k stars initially, which agrees quite well with the simulated ejection time of 
$\sim 4$ Gy shown in Figure~\ref{fig:fig1}.

To obtain Equation~\ref{eqn:strongtime}, we set:
\begin{equation}
\label{eqn:strongtime2}
\tau_{\rm si} = \frac{V_{\rm BH}}{\sqrt{3}\sigma_{\rm BH}\Sigma},
\end{equation}
where V$_{\rm BH} =$ (4/3)$\pi$(m/m$_{\rm BH}$)$^{3/2}$r$_{\rm h}^3$ is the volume within which all 
stellar-mass BHs in the cluster are confined after mass segregation \citep{heggie03} (which is approximately 
equal to one core radius in a typical GC; see below).  We express the 
BH velocity dispersion as $\sigma_{\rm BH} = \sigma_{\rm 0}\sqrt{m/m_{\rm BH}}$, where we have assumed 
energy equipartition and denote the central velocity dispersion as $\sigma_{\rm 0}$.\footnote{Note that the assumption of 
approximate energy equipartition here is only valid before the BHs decouple dynamically and undergo strong interactions with 
each other if the Spitzer instability is present.  This assumption is also valid whenever the number of BHs is very small, 
since in this limit the Spitzer instability does not occur.}  The gravitationally-focussed 
cross-section is $\Sigma = {\pi}$b$^2$, with the impact parameter b set to: 
\begin{equation}
\label{eqn:impact}
b^2 = \frac{2GMa_{\rm h}}{3\sigma_{\rm BH}^2},
\end{equation}
where M is the total mass of the IMBH-BH binary and a$_{\rm h}$ is the hard binary semi-major axis given 
in Equation~\ref{eqn:hard-soft}.  Here, we assume that the collisional cross-section is dominated by 
gravitational focusing.  Effectively, the strong interaction timescale given in Equation~\ref{eqn:strongtime} 
gives the time for the relevant volume within the cluster, namely the (approximate) volume of the core, to flow 
through the loss-cone.  


In Figure~\ref{fig:fig6} we show for several Milky Way GCs each of the half-mass relaxation timescale (filled 
circles), the single-binary encounter timescale (open squares) and the tidal capture timescale (crosses) as a 
function of the strong interaction timescale.  To calculate the strong interaction 
timescales, we assume an IMBH mass equal to 10$^3$ M$_{\rm \odot}$ in every cluster.  
The globular cluster sample and data are the same as used in \citet{leigh13c} (see Section 2 for a description of our 
sample), with the addition of M22 (NGC 6656).  All of these clusters are non-core-collapsed, and can be accurately fit 
by King models.  A summary of the clusters used in our sample along with the basic cluster parameters taken from 
\citet{harris96} and used to produce Figure~\ref{fig:fig6} can be found in Table~\ref{table:one}.  We also show 
in the last column of Table~\ref{table:one} the upper limit for the IMBH mass in each cluster, found by setting 
the sum of Equations~\ref{eqn:strongtime} and~\ref{eqn:t-rh} equal to 12 Gyr and solving for the IMBH mass.  Note 
that if the obtained upper limit on the IMBH mass is less than the mass of a typical stellar-mass BH (i.e. a few tens of 
M$_{\odot}$), then our results suggest that the detection of even a few BH binaries suggests that an IMBH does not exist 
in that cluster.

\begin{table*}
\caption{Properties of the GCs used in our sample, taken from \citet{harris96} and identified by their 
NGC number in Column 1.  Columns 2, 3 and 4 list, respectively, the cluster absolute V-band magnitudes, 
distance moduli and extinctions.  Columns 5 and 6 give the core and half-mass radius in arcminutes, respectively.  
Column 7 gives the logarithm of the central luminosity density in L$_{\odot}$ pc$^{-3}$.  Column 8 gives 
the core binary fraction taken from \citet{milone12}, or supplemented with the values calculated in \citet{leigh13c}.  
Columns 9 and 10 give the logarithms of the central and half-mass relaxation times, respectively, in years, taken 
directly from \citet{harris96}.  In Columns 11, 12 and 13 we provide the (logarithm of) strong interaction timescales, tidal 
capture timescales and single-binary exchange timescales in years, calculated using Equation~\ref{eqn:strongtime}, 
Equation~\ref{eqn:tc} and Equation~\ref{eqn:coll1+2}, respectively.  Finally, in Column 14, we provide the (logarithm of) 
the upper limit for the IMBH mass (in M$_{\odot}$), calculated by setting the sum of Equations~\ref{eqn:strongtime} 
and~\ref{eqn:t-rh} equal to 12 Gyr and solving for the IMBH mass.} 
\begin{tabular}{|c|c|c|c|c|c|c|c|c|c|c|c|c|c|}

\hline
ID  &    M$_{\rm V}$    &   (m-M)$_{\rm V}$   &    E(B-V)      &      r$_{\rm c}$    &     r$_{\rm h}$     &     log $\rho_{\rm 0}$    &    f$_{\rm b}$    &    log $\tau_{\rm rc}$       &  log $\tau_{\rm rh}$    &      log $\tau_{\rm si}$     &      log $\tau_{\rm tc}$      &     log $\tau_{\rm sb}$     &     log M$_{\rm IMBH}$    \\
(NGC)  &             &         &      &    (arcmin)    &     (arcmin)      &     (L$_{\odot}$ pc$^{-3}$)       &     &    (years)     &    (years)     &   (years)    &    (years)    &    (years)    &    (M$_{\odot}$)    \\
(1)  &     (2)    &   (3)   &         (4)       &          (5)         &          (6)           &     (7)     &        (8)       &           (9)        &   (10)     &    (11)     &     (12)     &   (13)     &     (14)    \\
\hline
 104  & -9.42  &  13.37  &  0.04   &    0.36  &  3.17   &   4.88  &    0.055   &  7.84  &  9.55  &   10.79  &   8.82  &  10.21  &   3.717   \\
1261  & -7.80  &  16.09  &  0.01   &    0.35  &  0.68   &   2.99  &    0.046   &  8.59  &  9.12  &   9.288  &  10.31  &  11.78  &   2.211   \\
2298  & -6.31  &  15.60  &  0.14   &    0.31  &  0.98   &   2.90  &    0.154   &  7.91  &  8.84  &   8.399  &  10.18  &  11.07  &   1.321   \\
3201  & -7.45  &  14.20  &  0.24   &    1.30  &  3.10   &   2.71  &    0.128   &  8.61  &  9.27  &    9.43  &  10.54  &  11.53  &   2.354   \\ 
4147  & -6.17  &  16.49  &  0.02   &    0.09  &  0.48   &   3.63  &    0.262   &  7.41  &  8.74  &   8.467  &  9.587  &  10.19  &   1.389   \\
4590  & -7.37  &  15.21  &  0.05   &    0.58  &  1.51   &   2.57  &    0.114   &  8.45  &  9.27  &    9.15  &  10.57  &  11.62  &   2.075   \\
5024  & -8.71  &  16.32  &  0.02   &    0.35  &  1.31   &   3.07  &    0.087   &  8.73  &  9.76  &    10.5  &  10.33  &   11.5  &   3.432   \\
5272  & -8.88  &  15.07  &  0.01   &    0.37  &  2.31   &   3.57  &    0.054   &  8.31  &  9.79  &    10.6  &  9.845  &  11.24  &   3.532   \\
5466  & -6.98  &  16.02  &  0.00   &    1.43  &  2.30   &   0.84  &    0.142   &  9.35  &  9.76  &   9.437  &  12.04  &  12.97  &   2.369   \\
5904  & -8.81  &  14.46  &  0.03   &    0.44  &  1.77   &   3.88  &     0.06   &  8.28  &  9.41  &   10.14  &  9.634  &  10.98  &   3.062   \\
5927  & -7.81  &  15.82  &  0.45   &    0.42  &  1.10   &   4.09  &    0.104   &  8.39  &  8.94  &   9.841  &  9.541  &  10.63  &   2.763   \\
5986  & -8.44  &  15.96  &  0.28   &    0.47  &  0.98   &   3.41  &    0.048   &  8.58  &  9.18  &   9.618  &  10.04  &  11.49  &   2.541   \\
6101  & -6.94  &  16.10  &  0.05   &    0.97  &  1.05   &   1.65  &      0.1   &  9.21  &  9.22  &   9.035  &  11.43  &  12.53  &   1.959   \\
6121  & -7.19  &  12.82  &  0.35   &    1.16  &  4.33   &   3.64  &    0.148   &  7.90  &  8.93  &   9.047  &  9.691  &   10.6  &   1.969   \\
6171  & -7.12  &  15.05  &  0.33   &    0.56  &  1.73   &   3.08  &    0.186   &  8.06  &  9.00  &   8.813  &  10.13  &  10.93  &   1.736   \\
6205  & -8.55  &  14.33  &  0.02   &    0.62  &  1.69   &   3.55  &     0.01   &  8.51  &  9.30  &    9.91  &  9.906  &  12.05  &   2.834   \\
6218  & -7.31  &  14.01  &  0.19   &    0.79  &  1.77   &   3.23  &    0.114   &  8.19  &  8.87  &   8.789  &  10.05  &  11.09  &   1.711   \\
6254  & -7.48  &  14.08  &  0.28   &    0.77  &  1.95   &   3.54  &    0.078   &  8.21  &  8.90  &   9.096  &  9.824  &  11.05  &   2.018   \\
6341  & -8.21  &  14.65  &  0.02   &    0.26  &  1.02   &   4.30  &    0.057   &  7.96  &  9.02  &   9.628  &  9.238  &  10.61  &   2.551   \\
6362  & -6.95  &  14.68  &  0.09   &    1.13  &  2.05   &   2.29  &     0.12   &  8.80  &  9.20  &   9.213  &  10.87  &  11.89  &   2.137   \\
6535  & -4.75  &  15.22  &  0.34   &    0.36  &  0.85   &   2.34  &    0.092   &  7.28  &  8.20  &   6.369  &  10.29  &  11.44  &  -0.710   \\
6584  & -7.69  &  15.96  &  0.10   &    0.26  &  0.73   &   3.33  &     0.09   &  8.13  &  9.02  &   9.012  &  9.951  &  11.11  &   1.935   \\
6637  & -7.64  &  15.28  &  0.18   &    0.33  &  0.84   &   3.84  &    0.124   &  8.15  &  8.82  &   9.157  &  9.631  &  10.63  &   2.079   \\
6652  & -6.66  &  15.28  &  0.09   &    0.10  &  0.48   &   4.48  &    0.344   &  7.05  &  8.39  &   8.167  &  8.974  &  9.407  &   1.088   \\
6656  & -8.50  &  13.60  &  0.34   &    1.33  &  3.36   &   3.63  &    0.046   &  8.53  &  9.23  &   9.853  &  9.869  &  11.34  &   2.777   \\
6723  & -7.83  &  14.84  &  0.05   &    0.83  &  1.53   &   2.79  &    0.062   &  8.79  &  9.24  &    9.52  &  10.52  &  11.85  &   2.444   \\
6779  & -7.41  &  15.68  &  0.26   &    0.44  &  1.10   &   3.28  &      0.1   &  8.33  &  9.01  &   9.225  &  10.05  &  11.16  &   2.148   \\
6838  & -5.61  &  13.80  &  0.25   &    0.63  &  1.67   &   2.83  &    0.304   &  7.54  &  8.43  &   7.343  &  10.18  &  10.69  &   0.265   \\ 
6934  & -7.45  &  16.28  &  0.10   &    0.22  &  0.69   &   3.44  &    0.092   &  8.20  &  9.04  &    9.27  &  9.889  &  11.04  &   2.193   \\
6981  & -7.04  &  16.31  &  0.05   &    0.46  &  0.93   &   2.38  &    0.098   &  8.72  &  9.23  &   9.252  &  10.78  &   11.9  &   2.176   \\
7089  & -9.03  &  15.50  &  0.06   &    0.32  &  1.06   &   4.00  &    0.094   &  8.48  &  9.40  &   10.37  &  9.641  &  10.78  &   3.296   \\
\hline
\end{tabular}
\label{table:one}
\end{table*}

To produce Figure~\ref{fig:fig6}, we multiplied the half-mass relaxation timescales taken from \citet{harris96} 
by the ratio of the average stellar mass in the core (assumed to be 0.5 M$_{\rm \odot}$ for all clusters) to the 
mass of the BH \citep{vishniac78} so that our estimates for the relaxation times are suitable to a 10 M$_{\odot}$ 
object (instead of an object with a mass equal to the average single star mass in the cluster core).  Similarly, both 
the single-binary 
encounter and tidal capture timescales are calculated for a single stellar-mass BH of mass 10 M$_{\rm \odot}$.  
For the encounter time, we take the average semi-major axis to be equal to 0.1 AU for all clusters.  This 
is because, we are primarily concerned with BH binaries that are actually observable, and the separation
must be small in order for BH-MS binaries to emit significantly in the X-ray, independent of the host
cluster properties (assuming all clusters are comparably old, as is the case for Mily Way GCs).  Most confirmed
stellar-mass BH binaries have separations slightly less than 0.1 AU \citep[e.g.][]{mcclintock06,kreidberg12}, however 
we do not account for any orbital decay which is likely 
for an initially eccentric orbit due to tidal dissipation at periastron.

As is clear from Figure~\ref{fig:fig6}, we confirm that the half-mass relaxation time is shorter than 
the strong interaction timescale in nearly all clusters.  More importantly, the sum of these timescales 
is typically much shorter than the cluster age, as is the case for $\sim 5/6$ of the GCs in our sample.  
It follows that, in these clusters, a sub-population of stellar-mass BHs should have sufficient time 
to not only mass segregate into the core, but also diffuse through the loss-cone of the IMBH-BH binary, assuming a central 
massive IMBH is present.  Therefore, \textit{in $\sim 83\%$ (i.e. 5/6) of the clusters considered here, the detection of 
even a single BH binary provides 
strong evidence against the presence of an IMBH.}  Our results suggest that this process should typically take on the order of 
a few Gyr, but it can approach and even exceed a Hubble time in the most massive MW GCs (in our sample these clusters are 
NGC 104, NGC 5024, NGC 5272, NGC 5904 and NGC 7089).  This is in rough agreement with the results of our $N$-body simulations, 
which suggest that a $\sim 10^5$ M$_{\odot}$ GC will have lost all its stellar-mass BHs within $\sim 4$ Gyr.  In fact, 
nearly all clusters in Figure~\ref{fig:fig6} with a total mass $\sim 10^5$ M$_{\odot}$ have a strong encounter 
time-scale on the order of a few Gyr, in rough agreement with the results of our $N$-body models.

Interestingly, 
the timescale for a stellar-mass BH to encounter a binary is longer than the tidal capture timescale in most GCs in 
our sample.  This implies that tidal captures should typically dominate the rate of BH binary formation.\footnote{We 
note that this conclusion is based on the \textit{present-day} binary fractions.  If the binary fractions were higher 
in the past, this could increase the significance of binary exchanges relative to tidal captures.}  However, provided 
the fraction of single-binary encounters that result in an exchange event is a significant fraction of the total 
number of such encounters, our results suggest that the difference is small, and hence the two mechanisms are of comparable 
significance.  


If, at a given time, the timescale for the formation of BH binaries is shorter than the strong interaction 
timescale, then a substantial population of BH binaries could form in the core before all BHs interact with the 
central IMBH.  If, on the other hand, the strong interaction timescale is shorter than the timescale 
for BH binary formation, then BHs should typically interact with the IMBH and be kicked out of the core before 
they acquire binary companions.  Whether or not a BH binary can co-exist in the cluster core with an IMBH and, 
if so, which formation mechanism is expected to dominate depends on the 
cluster in question, in particular its structural parameters, binary fraction and 
BH fraction.  The latter quantity is defined as the number of remaining BHs divided by 
the total number of objects in the cluster.  

\begin{figure}
\begin{center}
\includegraphics[width=\columnwidth]{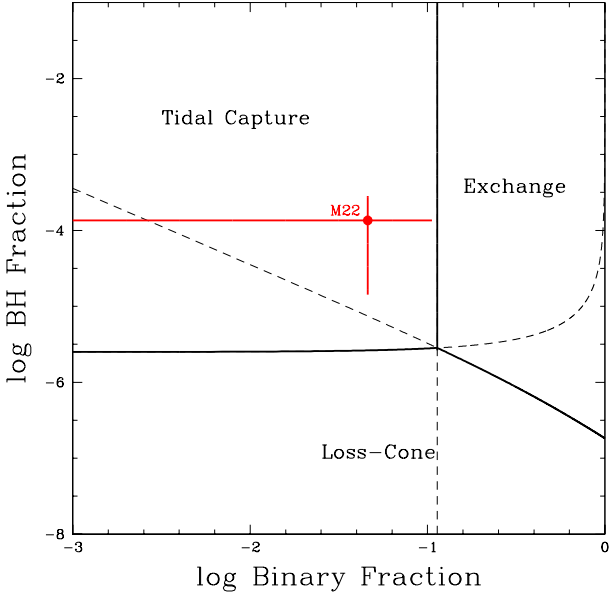}
\end{center}
\caption[The parameter space in the binary fraction-BH fraction plane for which at least some BHs 
could acquire a binary companion before being ejected from the cluster]{The binary fraction in 
the cluster core is plotted against the fraction of objects that are BHs.  This illustrates 
the parameter space for which the rate of BH depletion due to strong interactions with the 
IMBH-BH binary dominates over the rate of BH binary formation via either tidal capture of exchange interactions.  
This serves as a rough guide for the size of the BH population required to ensure that at least some BH binaries 
will form before undergoing a strong interaction with the IMBH-BH binary.  The BH and binary fractions are 
defined as the numbers of BHs and binaries, respectively, divided by the total number of cluster 
objects.  The dashed lines are obtained by equating the relevant timescales, namely those for strong interactions, 
tidal captures and binary exchanges.  The solid lines sector off the parameter space where each of 
these three processes dominate.  If the BH fraction is sufficiently low and strong interactions 
dominate, then the rate at which all BHs pass through the loss-cone exceeds the rate at which they acquire binary 
companions, and it is unlikely that any BH binaries should form before being ejected.  
The red point illustrates the observed estimates for the Galactic GC M22 
(NGC 6656), taken from \citet{strader12b} and \citet{milone12}.
\label{fig:fig7}}
\end{figure}

In Figure~\ref{fig:fig7}, we show \textit{for a given time} the regions of parameter space where either exchange 
interactions 
and/or tidal capture events (and hence the formation of a BH binary) should occur before all BHs in the core are ejected from 
the cluster, and vice versa.  
This illustrates the minimum BH fraction required in the core \textit{at a given time} for even one BH binary to form 
before the entire BH sub-population passes 
through the IMBH-BH binary loss-cone and are ejected from the cluster, in addition to the most likely 
BH binary formation mechanism.  If BHs segregate into the core at a rate that keeps the BH fraction in the core 
below this minimum value, then very few if any BH binaries should be present in the cluster at any given time.  
If observational constraints are available both for the BH and binary fractions 
in a given cluster core, 
a point can be placed in Figure~\ref{fig:fig7} for that cluster.  The dashed vertical line separates the tidal 
capture-dominated regime from the exchange-dominated regime -- if the point falls to the right (left) of this line, 
then exchanges (tidal captures) dominate.  


To construct Figure~\ref{fig:fig7} (including the parameter space where each mechanism dominates), we adopt the observed 
parameters for the Galactic GC M22 (NGC 6656), including a central luminosity density 
$\log \rho_{\rm 0} = 3.63$ \citep{harris96}, a core radius 1.7 pc \citep{harris96} and a core 
binary fraction f$_{\rm b} = 0.046 \pm 0.006$ \citep{milone12}.  Following \citet{strader12b}, 
we assume a population size of 5-100 stellar mass BHs in M22, all assumed to reside in the 
core, and an average BH mass m$_{\rm BH} =$ 10 M$_{\odot}$ \citep[e.g.][]{kalogera04}.  The 
central velocity dispersion is calculated from the virial theorem assuming a central 
mass-to-light ratio of 2, and an average stellar mass 0.5 M$_{\odot}$.  The average binary 
semi-major axis is chosen to be equal to the semi-major axis corresponding to the hard-soft 
boundary \citep[e.g.][]{heggie03}, since these binaries have the largest cross-sections for 
collision and are hence the most likely to undergo interactions with single BHs on the shortest 
timescales.  We note that this is likely an over-estimate for the formation rate of BH binaries 
via binary exchange encounters, since not all single-binary interactions will result in an 
exchange, nor will they necessarily produce a BH binary with a sufficiently short period to be 
observable as an X-ray binary.  Adopting a smaller average binary separation for BH binary 
formation via exchanges would increase the lower limit for the BH fraction needed for exchanges 
to dominate over BH depletion via strong interactions with the central IMBH-BH binary, and increase the binary 
fraction needed for exchanges to dominate over tidal capture. 

Figure~\ref{fig:fig7} illustrates that, if 5-100 stellar-mass BHs are present in the core of M22 at the 
present-day cluster age \citep{strader12b}, a handful of BHs should have had sufficient time to capture 
binary companions before being ejected from the cluster by the IMBH-BH binary, and could therefore be observable.  
But do we expect so many BHs to remain at an age of 12 Gyr if an IMBH 
is also present?  Assuming an IMBH mass of 10$^3$ M$_{\rm \odot}$, the answer is no, since the strong interaction 
timescale is less than a Hubble time, as shown by the red point in Figure~\ref{fig:fig6}.  In fact, the sum 
of the strong interaction and half-mass relaxation times is very nearly a Hubble time, 
yielding an upper limit on the mass of a possible IMBH of $\log$ (M$_{\rm IMBH}$/M$_{\odot}$) $\lesssim$ 2.8 
(see Table~\ref{table:one}).  
Thus, given that \citet{strader12b} estimate a population size of 5-100 stellar-mass BHs with 2 confirmed 
detections, these results argue against the presence of an IMBH in M22.  We emphasize that the 
detection of any additional BH binaries in M22 should further constrain the presence of an IMBH by 
lowering the upper limit on its mass.  Thus, the presence of even a handful of BH binaries in M22 
argues against the presence of an IMBH (with a non-negligible mass).
Independent of the presence of an IMBH, our results suggest that most BH binaries likely formed via 
tidal capture in M22.

To better illustrate the points raised above, we construct a toy Monte Carlo model to investigate the 
probability that a stellar-mass BH 
binary will remain in a globular cluster that contains an IMBH.  In our model,
10 M$_{\rm \odot}$ BHs form in a static globular cluster potential described by a 
Plummer model.  Initial cluster-centric radii for the BHs are drawn from a
Plummer density profile, and the BHs are then assumed to migrate inwards on a 
dynamical friction timescale \citep{binney87}.\footnote{We note that a Plummer density profile may not 
be realistic throughout the entire extent of the cluster, particularly near the cluster centre where 
the influence of the IMBH could steepen the density profile to better resemble a Bahcall-Wolf cusp.  However, 
this should not significantly affect our calculations for the dynamical friction timescale, which is dominated 
by the migration time at large clustercentric radii.  Regardless, adopting a cuspier density profile 
will only contribute to (slightly) decreasing the derived dynamical friction timescales.  Thus, the results of 
our toy Monte Carlo 
model can be regarded as upper limits.}  We run two models; one where all BHs are born 
single, and a second where all BHs are born in binaries.  

We use the same GC parameters for both models, chosen to be representative of a typical Milky 
Way globular cluster.  Specifically, we choose the cluster to have 10$^6$ total 
stars with a mean mass of 0.5 M$_{\odot}$, a binary fraction of 10\%, a half-mass radius of 3 pc,  
and a present-day age of 10 Gyr.  The resulting central density is about $2 \times 10^4$ pc$^{-3}$.  
We insert a $10^3$ M$_{\odot}$ IMBH with a 0.04 pc radius of influence in the center of the cluster.  
Then, we migrate the BHs inward towards the center of the cluster over 10 Gyr at a rate defined by the 
local dynamical friction timescale (updated at 1 Myr time-steps).  If a stellar-mass BH reaches the radius 
of influence of the IMBH before 10 Gyr, we assume that it is promptly ejected from the cluster or captured 
by the IMBH (and in either case would not be observed as an X-ray or radio source in the cluster).  Note 
that our results are insensitive to the precise distance used, and our answer is more or less unaffected 
if we adopt the hard binary separation given in Equation~\ref{eqn:hard-soft} instead of the radius of 
influence.

For the model where all BHs are born single, we check if a given BH will obtain a companion along its 
inward migration either through a binary-single exchange encounter or a tidal capture event, 
following the timescales given in \citet{kalogera04}, 
and accounting for the binary fraction in the tidal capture timescale as 
mentioned above.  We set the exchange encounter radius to 1 AU, and the tidal capture radius to 
5 R$_{\odot}$.  We take a running average of these respective timescales for a 
given BH as it migrates towards the cluster core.  To account for the stochastic 
nature of stellar encounters in star clusters, at each time step we draw a 
random number between 0 and 1 and if the encounter timescale multiplied 
by this random number is less than the current model time we assume the BH gained a binary 
companion (following \citet{ivanova05b}).  Certainly a more detailed model is desirable (for instance as in 
\citet{morscher13}), but is beyond the scope of this paper.

\begin{figure}
\begin{center}
\includegraphics[width=\columnwidth]{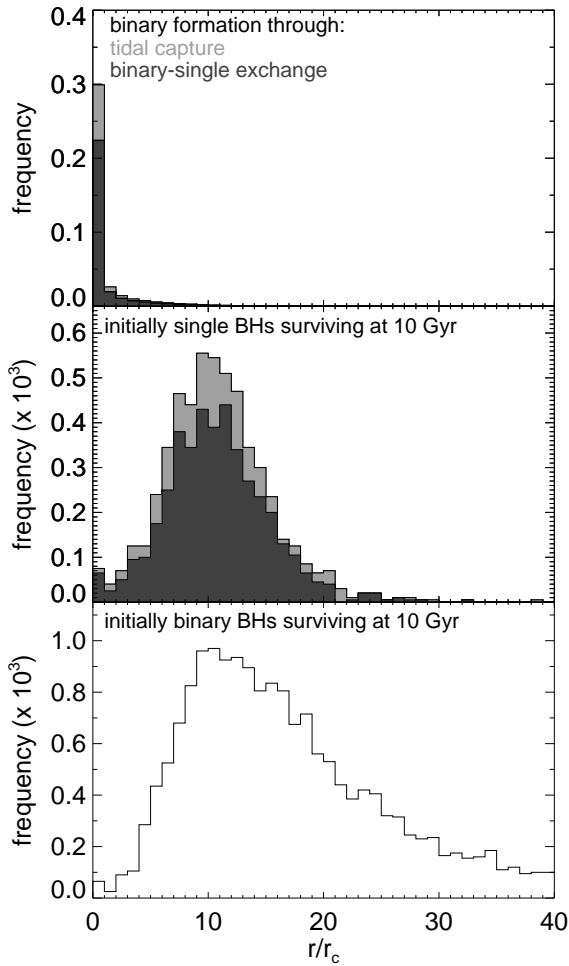}
\end{center}
\caption[Results from our toy Monte Carlo model for the radial migration of BHs in GCs]{Results from our toy 
Monte Carlo model in which 10 M$_{\odot}$ BHs are born within a Plummer 
density distribution and allowed to migrate inwards toward the cluster core on 
a dynamical friction timescale, over a cluster lifetime of 10 Gyr.  The cluster hosts a 10$^3$ M$_{\odot}$ IMBH 
at its centre.  In both insets, we plot stacked histograms and 
designate between BHs that obtain companions through tidal capture and exchange events via the light- and 
dark-grey bins, respectively.  The top inset shows the clustercentric radii at which isolated BHs obtain
companions at any time during the inward migration, the vast majority of which obtain companions within 1
core radius from the cluster centre.  The middle and bottom insets show the frequencies of BH binaries that 
remain in the cluster at 10 Gyr, assumed either to be born single (middle inset) or in a binary (bottom inset) 
at the indicated clustercentric radius.
\label{fig:fig8}}
\end{figure}

Nonetheless, the results from our toy model are instructive, and are shown in 
Figure~\ref{fig:fig8}.  First, we note that the vast majority 
of the single BHs that obtain companions do so within one core radius (r$_{\rm c}$) from the cluster centre, 
where the density is high and therefore the encounter timescales are low.  Given our assumptions about 
the cluster structure and binary frequency, exchange encounters are the most likely mechanism for 
initially single BHs to obtain companions.  Interestingly, of those BH binaries expected to remain in the 
cluster, our 
model predicts that they will most likely be found at radii of $\sim$ 10 r$_{\rm c}$ or beyond.  
These BHs were born significantly further out from the center, at radii of around 20 r$_{\rm c}$, where the 
dynamical friction timescale is long, and therefore these BHs do not have time to migrate fully into the core.  

Given a typical GC hosting an IMBH, our model also predicts that there is only about a 0.5\% (2\%) conditional 
probability that a stellar-mass BH that forms single (in a binary) will remain in a binary within a typical 
globular cluster at the present day (which we call P(A)), given that this globular cluster has an IMBH (wich we 
call P(B)).  If we assume that BHs are equally likely to form single as they are to form in binaries, then this 
conditional probability, P(A$|$B), equals 1.25\%.  To obtain these probabilities, we ran 2 $\times$ 10$^5$ BHs 
through our Monte Carlo model for each case (i.e. all BHs are born single and all BHs are born in binaries), and counted 
the number of BHs that exist in a BH in a binary within the cluster at 10 Gyr.  The results are plotted 
in Figure~\ref{fig:fig8} in the form of histograms, which are then integrated to obtain the total probabilities 
listed above.

Thus, to first-order, these results suggest that, of all stellar-mass BHs born in the cluster, on the order of 
$\sim$ 1\% will still remain in the cluster by avoiding a direct encounter with the central IMBH, and also have 
a binary companion at a cluster age of 10 Gyr.  This estimate could increase, depending on how many BHs remain 
bound to the cluster and also retain their binary companion after undergoing one or more interactions with 
the IMBH.  Additionally, note that if P(A) and P(B) are known (e.g., through future observational efforts), one 
can use Bayes' Theorem with such a model to calculate P(B$|$A) = P(A$|$B)P(B)/P(A), the probability that a GC 
has an IMBH given the detection of a stellar-mass BH in a binary.


We caution that the results presented in this section are sensitive to the initial cluster conditions, 
which are poorly known.  This includes the initial stellar mass function at the high-mass end, the 
initial-final mass relation for BHs, the magnitude and frequency of natal kicks, the initial cluster
mass and structural properties, etc.  An interesting example that relates to the issue of BH retention
is the influence of gas damping during the gas-embedded phase of cluster evolution.  In particular,
gas damping can significantly accelerate the rate of dynamical decoupling of a massive sub-population
such as BHs, and even prevents any stable energy equilibrium from occurring between different mass species, even 
if Spitzer's Criterion is 
satisfied.  Thus, gas damping can cause BHs to be ejected on shorter timescales, or even to be
ejected when they otherwise would not be \citep{leigh13a}.

\subsection{Will a BH binary survive a direct encounter with an IMBH-BH binary?} \label{direct}

In this section, we obtain a rough estimate for an upper limit for the BH binary survival probability during 
a strong interaction with an IMBH-BH binary, and defer more detailed numerical scattering experiments 
\citep[e.g.][]{colpi03} to a future paper.

Consider two limiting cases.  In the first scenario, the semi-major 
axis of the IMBH-BH binary is comparable to or smaller than that 
of the BH-MS binary.  In this case, the disruption time for the BH-MS binary is given by
Equation~\ref{eqn:strongtime}, since the binding energy of the BH-MS binary is sufficiently low (in terms of
its absolute value) compared to that of the IMBH-BH binary that a direct encounter will almost
certainly disrupt the BH-MS binary.  In the second scenario, the semi-major axis of the IMBH-BH 
binary is much larger than that of the BH-MS binary.  In this case, the BH-MS binary could survive 
a direct encounter with the IMBH-BH binary, provided its distance of closest approach to the IMBH exceeds 
the tidal disruption radius for the BH-MS binary.  The tidal disruption radius for the BH-MS binary is 
r$_{\rm t} \sim$ (M$_{\rm BH-MS}$/M$_{\rm IMBH}$)$^{1/3}$a, where M$_{\rm IMBH}$ is the IMBH mass, M$_{\rm BH-MS}$ 
is the mass of the BH-MS binary and a is its semi-major axis.  

Based on the above, an order-of-magnitude estimate for the maximum fraction of strong interactions between a BH-MS binary and 
IMBH-BH binary (with a semi-major axis given in Equation~\ref{eqn:hard-soft}) that leave the BH-MS 
binary intact can be obtained.  Our estimate for the survival probability is an upper limit, and we come back to this 
below.  It comes from the ratio of the gravitationally-focused cross-sections, or:
\begin{equation}
\label{eqn:fsurvive}
f_{\rm surv} = 1 - \sqrt{\frac{(M_{\rm IMBH}+M_{\rm BH})r_{\rm t}}{M_{\rm IMBH}a_{\rm h}}}.
\end{equation}
Here, we adopt for the IMBH-BH binary a semi-major axis given in Equation~\ref{eqn:hard-soft}, which corresponds to the point at which the
main mechanism driving the orbital decay changes from dynamical friction to scattering interactions.  For example, 
assuming M$_{\rm IMBH} =$ 10$^3$ M$_{\rm \odot}$, Equation~\ref{eqn:fsurvive} yields a survival fraction of $\sim 65-95$\% for 
the sample of Galactic GCs considered in the previous section.  

As stated, our quoted estimate for the survival fraction of BH binaries after direct encounters with the IMBH-BH binary 
represents an upper limit.  We do not consider the \textit{rate} of 
orbital decay, and hence the time the IMBH-BH binary spends at each semi-major axis.  The rate of orbital decay reaches a minimum just 
beyond a$_{\rm GW}$, since here GW emission is not yet effective and the rate of scattering interactions is low since the 
cross-section of the IMBH-BH binary is small.  In general, as the orbital separation of the IMBH-BH binary falls 
below a$_{\rm h}$ due to scattering interactions, the survival fraction decreases.  This continues until the semi-major 
axis of the IMBH-BH binary reaches that of the BH-MS binary (i.e. $\sim 0.1 AU$), at which point the survival probability 
drops to approximately zero.  Additionally, our estimate ignores resonant interactions which 
increase the number of close approaches that can occur between the BH-MS binary and the IMBH, and
hence the probability of disruption.


\section{Summary and Discussion} \label{summary}

In this paper, we use a combination of $N$-body simulations and analytic methods to address the fate of a 
sub-population of stellar-mass BHs in clusters 
hosting a central IMBH.  We further consider the possible presence of BHs with binary companions, 
weighed against the ejecting/disrupting influence of the IMBH.  Our main results can be summarized as follows:

\begin{itemize}
\item A central IMBH will typically capture the most massive stellar-mass BH in a binary early 
on in the cluster lifetime.  The IMBH will retain a BH companion until the BH spirals in, due to scattering 
interactions with neighbouring objects and gravitational wave emission, and merges.  
\item For typical IMBH masses in GCs following from the M-$\sigma$ relation 
\citep{kruijssen13}, gravitational wave emission is likely to play an important role in deciding 
the evolution of the IMBH-BH binary orbital parameters, often while hardening interactions 
with the surrounding stellar population also drive orbital decay.  Thus, GW emission should be 
accounted for in future related studies.
\item \textit{In most Galactic GCs, the detection of one or more BH binaries can be used to place strong constraints on the 
possible presence of an IMBH.}\footnote{It is important to make the distinction, however, that the absence of 
detected BH binaries does not automatically mean that an IMBH is present.}
\item If the (instantaneous) timescale for BHs in the cluster core to flow through the loss-cone and undergo 
strong interactions with the central IMBH-BH binary is longer than the formation time for BH binaries, then at 
least some isolated 
BHs will have time to acquire binary companions \textit{before} interacting with the IMBH, at which 
point they are typically ejected from the cluster.
\end{itemize}

Importantly, we assume throughout this paper that the IMBH mass is $\sim 1$\% of the \textit{present-day} 
cluster mass.  This is larger than predicted by extending the M-$\sigma$ relation observed for SMBHs in galactic 
nuclei \citep[e.g.][]{ferrarese00,kruijssen13}, but not unreasonable if GCs were once considerably more 
massive than their present-day masses suggest.  For lower IMBH masses, 
even the IMBH can be ejected from the cluster due to dynamical interactions with other massive remnants.  In this 
case, our results do not apply.  If an 
IMBH is retained, our results for each initial total cluster mass can be extended to lower IMBH masses as follows.  
First, for a given number of BHs (and host cluster properties), lowering the central IMBH mass should increase the 
timescale for all BHs to 
be ejected from the cluster.  This is because individual interactions between the IMBH and BHs tend to be less 
energetic, and thus less likely to accelerate BHs to velocities that exceed the escape speed.  Second, 
as shown in Figure~\ref{fig:fig4}, for 
a given cluster mass and hence central velocity dispersion, lowering the central IMBH mass in turn increases 
the critical semi-major axis a$_{\rm h}$ at which hardening interactions begin to dominate the rate of orbital decay, 
and decreases the critical semi-major axis a$_{\rm GW}$ at which gravitational wave emission begins to dominate.  
Thus, for lower IMBH masses, the role of hardening interactions increases at small semi-major axes, where the IMBH-BH 
binary cross-section is at its smallest.  This prolongs the lifetime of the IMBH-BH binary by increasing the 
time it spends with a semi-major axis $\gtrsim$ a$_{\rm GW}$.  The net result is an increase in the merger time 
of the IMBH-BH binary.  The evolution of the IMBH-BH binary (along with the time for it to merge) can be 
accurately modeled with minimal computational expense for semi-major axes a $<$ a$_{\rm h}$ using a combination of 
analytic methods and numerical scattering experiments.  This can be done for any combination of IMBH-BH binary orbital 
parameters, as well as central cluster densities 
and velocity dispersions \citep{quinlan96,sesana06}.  

We caution that the timescales presented in this paper depend to some degree on the initial cluster 
conditions, which are in general quite uncertain.  This includes the initial density and concentration, the initial 
cluster mass, the degree to which cluster's are initially tidally over- or under-filling, the initial numbers and 
spatial positions of stellar-mass BHs and binaries, the maximum BH mass, etc.  We also do not consider BHs formed 
through binary evolution due to the considerable uncertainties inherent to this formation pathway, although we expect 
this contribution to the total number of BHs to be small compared to 
the number of primordial BHs.  All of these uncertainties 
affect our estimates for the initial numbers, ejection times and survival probabilities of BHs, as well as the characteristic 
semi-major axes of any central IMBH-BH binary (i.e. a$_{\rm h}$ and a$_{\rm GW}$) and thus its quantitative 
evolution.  This paper illustrates that the numbers of BH binaries in GCs can potentially be used to constrain 
the possible presence of an IMBH (or at least an upper limit for its mass), and provides a useful benchmark 
for future studies focusing on individual clusters.

Interestingly, if an IMBH is present \textit{and} it has a BH binary companion at the present-day cluster 
age, the distribution of ejected stars and/or compact binaries could be asymmetric.  This is because 
the escapers are typically ejected in the plane of the IMBH-BH binary, but forming a wide jet if the binary 
is eccentric.  This raises 
the interesting possibility of observing a weakly visible and wide jet of escaping stars and/or 
compact binaries.  Indeed, such an asymmetry may already have been observed.  For example, 
\citet{grindlay06} reported a possible asymmetry in the spatial distribution of binaries (emitting X-rays) 
about the cluster centre in the core-collapsed globular cluster NGC 6397.  
The issue of whether or not such a wide jet would actually be observable in some clusters will be 
the topic of a future paper.  However, we note that, to the best of our knowledge, such an 
asymmetry in the distribution of ejected objects can only be produced via a 
massive central \textit{binary}.  This is because the distribution of ejection velocities should 
be isotropic if they originated from a strong interaction with an isolated IMBH, or even during a resonant 
encounter with a stellar-mass BH or even NS binary \citep[e.g.][]{drukier03}.  Thus, an isotropic distribution 
of high-velocity escapers cannot be used to unambiguously identify the presence of an IMBH, since 
encounters with BH or NS binaries could also reproduce such a signature.  Similarly, previous 
authors have argued that, although the kinematic signatures will be difficult 
to measure with present-day telescopes, an IMBH-IMBH binary could produce a population of a few hundred 
suprathermal stars moving at anomalously high velocities \citep{mapelli05}.  The orbital angular momenta 
of these objects 
should be preferentially aligned with that of the IMBH-BH binary, creating an anisotropy in the 
angular momentum distribution of cluster stars.  Thus far, no asymmetry is clearly present in either NGC 104 
\citep{mclaughlin06} or $\omega$ Cen \citep{anderson10}, although a signal may be present in NGC 2808 
\citep{luetzgendorf12}.  Our results suggest that, in clusters with strong interaction 
times that are longer than the BH binary formation time, such an anisotropy may also be 
present (and easier to observe), and should be looked for, in the distribution of X-ray binaries in the cluster.  


Our results can be extended to include as well neutron star X-ray binaries, although more work is needed in 
this direction.  As shown in Figure~\ref{fig:fig1}, it takes longer 
for a cluster's NS population to be depleted via interactions with a central IMBH than it does 
for its BH population.  This is due to the smaller masses of NSs, and hence their longer 
relaxation times.  This raises the interesting possibility of using the ratio of BH to NS X-ray 
binaries (or simply the absolute numbers of NS X-ray binaries in clusters with very few or no BHs) as a 
diagnostic for probing the possible presence of an IMBH, since it 
accelerates the rate of ejection of (preferentially massive) objects from the cluster.  In massive clusters 
with very long relaxation times, it follows that the BH X-ray binary population should be more depleted 
due to strong interactions with the IMBH than the NS X-ray binaries.  Moreoever, NS X-ray binaries could 
potentially provide a more useful 
tracer of an anisotropy in the angular momentum distribution in clusters for which the timescale
for all BHs to be ejected is shorter than the cluster age.  

%


\section*{Acknowledgments}

We would like to thank Elena Rossi, Mirek Giersz, Coleman Miller and Steve Zepf for useful discussions and suggestions.  
N. W. C. L. is grateful for the generous support of an NSERC Postdoctoral Fellowship.  N. L. thanks Holger Baumgardt for 
his considerable assistance and guidance, in particular helping to implement the collision procedures in NBODY6.  
A. M. G. is funded by 
a National Science Foundation Astronomy and Astrophysics Postdoctoral Fellowship under Award No. AST-1302765.  
C. O. H. is funded by an NSERC Discovery Grant and an Ingenuity New Faculty Award.


\bsp

\label{lastpage}


\begin{thebibliography}{99}

\bibitem[\protect\citeauthoryear{Aarseth}{1999}]{aarseth99} Aarseth S. J. 1999, 
PASJ, 111, 1333
\bibitem[\protect\citeauthoryear{Anderson \& van der Marel}{2010}]{anderson10} Anderson J., 
van der Marel R. P. 2010, ApJ, 710, 1032
\bibitem[\protect\citeauthoryear{Ayasli \& Joss}{1982}]{ayasli82} Ayasli S., Joss P. C. 1982, 
ApJ, 256, 637
\bibitem[\protect\citeauthoryear{Bahcall \& Ostriker}{1975}]{bahcall75} Bahcall J. N., 
Ostriker J. P. 1975, Nature, 256, 23
\bibitem[\protect\citeauthoryear{Baumgardt et al.}{2003}]{baumgardt03} Baumgardt H., Piet H., 
Makino J., McMillan S., Porteies Zwart S. 2003, ApJL, 582, L21
\bibitem[\protect\citeauthoryear{Binney \& Tremaine}{1987}]{binney87}
  Binney J., Tremaine S., 1987, Galactic Dynamics (Princeton:
  Princeton University Press)
\bibitem[\protect\citeauthoryear{Bellazzini et al.}{2008}]{bellazzini08} Bellazzini R. A., 
Ibata R. A., Chapman S. C., Mackey A. D., Monaco L., Irwin M. J., Martin N. F., 
Lewis G. F., Dalessandro E. 2008, AJ, 136, 1147
\bibitem[\protect\citeauthoryear{Breen \& Heggie}{2013}]{breen13} Breen P. G., Heggie D. C. 
2013, MNRAS, 432, 2779
\bibitem[\protect\citeauthoryear{Chomiuk et al.}{2013}]{chomiuk13} Chomiuk L., Strader J., 
Maccarone T. J., Miller-Jones J. C. A., Heinke C., Noyola E., Seth A. C., Ransom S. 2013, 
ApJ, 777, 69
\bibitem[\protect\citeauthoryear{Clark}{1975}]{clark75} Clark F. O. 1975, ApJL, 200, L115
\bibitem[\protect\citeauthoryear{Colpi, Possenti \& Gualandris}{2002}]{colpi02} Colpi M., 
Possenti A., Gualandris A. 2002, ApJL, 570, L85
\bibitem[\protect\citeauthoryear{Colpi, Mapelli \& Possenti}{2003}]{colpi03} Colpi M., 
Mapelli M., Possenti A. 2003, ApJ, 599, 1260
\bibitem[\protect\citeauthoryear{Downing et al.}{2010}]{downing10} Downing J. M. B.,
Benacquista M. J., Giersz M., Spurzem R. 2010, MNRAS, 407, 1946
\bibitem[\protect\citeauthoryear{Drukier \& Bailyn}{2003}]{drukier03} Drukier G. A., Bailyn C. D. 2003, 
ApJ, 597, L125
\bibitem[\protect\citeauthoryear{Duquennoy \& Mayor}{1991}]{duquennoy91} Duquennoy A.,
Mayor M. 1991, A\&A, 248, 485
\bibitem[\protect\citeauthoryear{Ferrarese \& Merritt}{2000}]{ferrarese00} Ferrarese L., 
Merritt D. 2000, ApJL, 539, L9
\bibitem[\protect\citeauthoryear{Fryer \& Kalogera}{2001}]{fryer01} Fryer C. L.,
Kalogera V. 2001, ApJ, 554, 548
\bibitem[\protect\citeauthoryear{Fryer et al.}{2012}]{fryer12} Fryer C. L.,
Belczynski K., Wiktorowicz G., Dominik M., Kalogera V., Holz D. E. 2012, ApJ, 749, 91
\bibitem[\protect\citeauthoryear{Fregeau et al.}{2004}]{fregeau04} Fregeau J. M., 
Cheung P., Portegies Zwart S. F., Rasio F. A. 2004, MNRAS, 352, 1
\bibitem[\protect\citeauthoryear{Gebhardt, Rich \& Ho}{2005}]{gebhardt05} Gebhardt K., 
Rich R. M., Ho L. C. 2005, ApJ, 634, 1093
\bibitem[\protect\citeauthoryear{Gerssen et al.}{2002}]{gerssen02} Gerssen J., van der Marel R. P.,
Bebhardt K., Guhathakurta P., Peterson R. C., Pryor C. 2002, AJ, 124, 3270
\bibitem[\protect\citeauthoryear{Gill et al.}{2008}]{gill08} Gill M., Trenti M., Miller M. C., 
van der Marel R., Hamilton D., Stiavelli M. 2008, ApJ, 686, 303
\bibitem[\protect\citeauthoryear{Grindlay et al.}{1976}]{grindlay76} Grindlay J., Gursky H., 
Schnopper H., Parsignault D. R., Heise J., Brinkman A. C., Schrijver J. 1976, ApJL, 205, L127
\bibitem[\protect\citeauthoryear{Grindlay}{2006}]{grindlay06} Grindlay J. E. 2006, 
Advances in Space Research, 38, 2923
\bibitem[\protect\citeauthoryear{Haggard et al.}{2013}]{haggard13} Haggard D., Cool A. M., 
Heinke C. O., van der Marel R., Cohn H. N., Lugger P. M., Anderson J. 2013, ApJ, 773, 31
\bibitem[\protect\citeauthoryear{Harris}{1996, 2010 update}]{harris96}
  Harris, W. E. 1996, AJ, 112, 1487 (2010 update)
\bibitem[\protect\citeauthoryear{Heggie \& Hut}{2003}]{heggie03}
  Heggie D. C., Hut P. 2003, The Gravitational Million-Body Problem:
  A Multidisciplinary Approach to Star Cluster Dynamics (Cambridge:
  Cambridge University Press)
\bibitem[\protect\citeauthoryear{Heggie}{1974}]{heggie74} Heggie D. C. 1974, 
Celestial Mechanics, 10, 217
\bibitem[\protect\citeauthoryear{Heggie}{1975}]{heggie75} Heggie
  D. C. 1975, MNRAS, 173, 729
\bibitem[\protect\citeauthoryear{Heggie \& Giersz}{2013}]{heggie13} Heggie D. C., 
Giersz M. 2013, MNRAS, accepted (arXiv:1401.3657)
\bibitem[\protect\citeauthoryear{Ibata et al.}{2009}]{ibata09} Ibata R., Bellazzini M.,
Chapman S. C., Dalessandro E., Ferraro F., Irwin M., Lanzoni B., Lewis G. F., Mackey A. D.,
Miocchi P., Varghese A. ApJ, 699, L169
\bibitem[\protect\citeauthoryear{Illingworth \& King}{1977}]{illingworth77} Illingworth G. D., 
King I. R. 1977, ApJL, 218, L109
\bibitem[\protect\citeauthoryear{Ivanova et al.}{2005a}]{ivanova05a} Ivanova N., Rasio F. A.,
Lombardi J. C. Jr., Dooley K. L., Proulx Z. F. 2005, ApJL, 621, L109
\bibitem[\protect\citeauthoryear{Ivanova et al.}{2005b}]{ivanova05b} Ivanova N., Belczynski K., 
Fregeau J. M., Rasio F. A. 2005, MNRAS, 358, 572 
\bibitem[\protect\citeauthoryear{Kalogera, King \& Rasio}{2004}]{kalogera04} Kalogera V., 
King A. R., Rasio F. A. 2004, ApJL, 601, L171
\bibitem[\protect\citeauthoryear{Kroupa}{2002}]{kroupa02} Kroupa P. 2002, Science, 295, 82
\bibitem[\protect\citeauthoryear{Kroupa}{2008}]{kroupa08} Kroupa P. 2008, Cambridge $N$-body Lectures,
Lecture Notes in Physics, 760. Springer-Verlag, Berlin, p. 181
\bibitem[\protect\citeauthoryear{Kroupa et al.}{2013}]{kroupa13} Kroupa P., Weidner C.,
Pflamm-Altenburg J., Thies I., Dabringhausen J., Marks M., Maschberger T. 2013,
Planets, Stars and Stellar Systems Vol. 5, ed. Oswalt T. D. \& Gilmore G.
(Springer Science \& Business Media Dordrecht:  Dordrecht), 115-242
\bibitem[\protect\citeauthoryear{Kreidberg et al.}{2012}]{kreidberg12} Kreidberg L., Bailyn C., 
Farr W. M., Kalogera V. 2012, ApJ, 757, 36
\bibitem[\protect\citeauthoryear{Kruijssen \& L\"utzgendorf}{2013}]{kruijssen13} Kruijssen J. M. D., 
L\"utzgendorf N. 2013, MNRAS Letters, 434, L41
\bibitem[\protect\citeauthoryear{Kulkarni, Hut \& McMillan}{1993}]{kulkarni93} Kulkarni S. R., Hut P., 
McMillan S. 1993, Nature, 364, 421
\bibitem[\protect\citeauthoryear{Larson}{1984}]{larson84} Larson R. B. 1984, MNRAS, 210, 763
\bibitem[\protect\citeauthoryear{Leigh \& Sills}{2011a}]{leigh11a} Leigh
  N. W., Sills A. 2011, MNRAS, 410, 2370 (Leigh \& Sills 2011a)
\bibitem[\protect\citeauthoryear{Leigh, Sills \& Knigge}{2011b}]{leigh11b} Leigh N. W.,
Sills A., Knigge C. 2011, MNRAS, 415, 377 (Leigh, Sills \& Knigge 2011b)
\bibitem[\protect\citeauthoryear{Leigh et al.}{2013a}]{leigh13a} Leigh N. W., 
Mastrobuono-Battisti A., Perets H. B., B\"oker T. 2013, MNRAS, 441, 919 (Leigh et al. 2013a)
\bibitem[\protect\citeauthoryear{Leigh et al.}{2013b}]{leigh13b} Leigh N. W. C., 
B\"oker T., Maccarone T. J., Perets H. B. 2013, MNRAS, 429, 2997 (Leigh et al. 2013b)
\bibitem[\protect\citeauthoryear{Leigh et al.}{2013c}]{leigh13c} Leigh N. W. C., Knigge C., 
Sills A., Perets H. B., Sarajedini A., Glebbeek E. 2013, MNRAS, 428, 897 (Leigh et al. 2013c)
\bibitem[\protect\citeauthoryear{Leonard}{1989}]{leonard89} Leonard P. J. T. 1989, AJ, 98, 217 
\bibitem[\protect\citeauthoryear{L\"utzgendorf et al.}{2012}]{luetzgendorf12} L\"utzgendorf N., Gualandris A., 
Kissler-Patig M., Gebhardt K., Baumgardt H., Noyola E., Kruijssen J. M. D., Jalali B., de Zeeuw P. T., 
Neumayer N. 2012, A\&A, 543, 82
\bibitem[\protect\citeauthoryear{L\"utzgendorf et al.}{2013}]{luetzgendorf13a} L\"utzgendorf N., 
Kissler-Patig M., Gebhardt K., Baumgardt H., Noyola E., de Zeeuw P. T., Neumayer N., Jalali B., 
Feldmeier A. 2013, A\&A, 552, 49
\bibitem[\protect\citeauthoryear{L\"utzgendorf, Baumgardt \& Kruijssen}{2013}]{luetzgendorf13b} L\"utzgendorf N., 
Baumgardt H., Kruijssen J. M. D. 2013, A\&A, 558, 117
\bibitem[\protect\citeauthoryear{Maccarone et al.}{2007}]{maccarone07}
Maccarone T. J., Kundu A., Zepf S. E., Rhode K. L. 2007, Nature, 445, 183
\bibitem[\protect\citeauthoryear{Maccarone \& Servillat}{2008}]{maccarone08} Maccarone T. J., 
Servillat M. 2008, MNRAS, 389, 379 
\bibitem[\protect\citeauthoryear{Maccarone et al.}{2011}]{maccarone11} Maccarone T. J., 
Kundu A., Zepf S. E., Rhode K. L. 2011, MNRAS, 410, 1655
\bibitem[\protect\citeauthoryear{Mapelli et al.}{2005}]{mapelli05} Mapelli M., Colpi M., 
Possenti A., Sigurdsson S. 2005, MNRAS, 364, 1315
\bibitem[\protect\citeauthoryear{McLaughlin et al.}{2006}]{mclaughlin06} McLaughlin D. E., Anderson J., 
Meylan G., Gebhardt K., Pryor C., Minniti D., Phinney S. 2006, ApJS, 166, 37 
\bibitem[\protect\citeauthoryear{McClintock \& Remillard}{2006}]{mcclintock06} McClintock J. E., 
Remillard R. A. 2006, in Compact stellar X-ray sources, Cambridge Astrophysics Series 39, ed. W. Lewin 
\& M. van der Klis (Cambridge:  Cambridge University Press), 157
\bibitem[\protect\citeauthoryear{Merritt}{2013}]{merritt13} Merritt D. 2013, Dynamics and Evolution 
of Galactic Nuclei (Princeton:  Princeton University Press)
\bibitem[\protect\citeauthoryear{Mikkola}{1985}]{mikkola85} Mikkola S. 1985, MNRAS, 215, 171
\bibitem[\protect\citeauthoryear{Milone et al.}{2012}]{milone12}
  Milone A. P., Piotto G., Bedin L. R., Aparicio A., Anderson J.,
  Sarajedini A., Marino A. F., Moretti A., Davies M. B., Chaboyer B., Dotter A.,
  Hempel M., Marin-Franch A., Majewski S., Paust N. E. Q., Reid I. N.,
  Rosenberg A., Siegel M. 2012, A\&A, 540, 16
\bibitem[\protect\citeauthoryear{Milosavljevic \& Merritt}{2001}]{milosavljevic01} Milosavljevic M., 
Merritt D. 2001, ApJ, 563, 34
\bibitem[\protect\citeauthoryear{Moody \& Sigurdsson}{2009}]{moody09} Moody K., 
Sigurdsson S. 2009, ApJ, 690, 1370
\bibitem[\protect\citeauthoryear{Morscher et al.}{2013}]{morscher13} Morscher M., Umbreit S., 
Farr W. M., Rasio F. A. 2013, ApJL, 763, L15
\bibitem[\protect\citeauthoryear{Newell, Da Cost \& Norris}{1976}]{newell76} Newell B., Da Costa G. S., 
Norris J. 1976, ApJL, 208, L55
\bibitem[\protect\citeauthoryear{Noyola, Gebhardt \& Bergmann}{2008}]{noyola08} Noyola E., Gebhardt K., 
Bergmann M. 2008, ApJ, 676, 1008
\bibitem[\protect\citeauthoryear{Pasquato et al.}{2009}]{pasquato09} Pasquato M., 
Trenti M., De Marchi G., Gill M., Hamilton D. P., Miller M. C., Stiavelli M., van der Marel R. P. 
2009, ApJ, 699, 1511
\bibitem[\protect\citeauthoryear{Portegies Zwart \& McMillan}{2000}]{portegieszwart00} Portegies Zwart S. F., 
McMillan S. L. W. 2000, ApJL, 528, L17
\bibitem[\protect\citeauthoryear{Portegies Zwart et al.}{2004}]{portegieszwart04} Portegies Zwart S. F., 
Baumgardt H., Hut P., Makino J., McMillan S. L. W. 2004, Nature, 428, 724 
\bibitem[\protect\citeauthoryear{Quinlan \& Shapiro}{1990}]{quinlan90} Quinlan G. D.,
Shapiro S. L. 1990, ApJ, 356, 483
\bibitem[\protect\citeauthoryear{Quinlan}{1996}]{quinlan96} Quinlan G. D. 1996, New Astronomy, 1, 35
\bibitem[\protect\citeauthoryear{Sana, Gosset \& Evans}{2009}]{sana09} Sana H., Gosset E., 
Evans C. J. 2009, MNRAS, 400, 1479
\bibitem[\protect\citeauthoryear{Sana, James \& Gosset}{2011}]{sana11} Sana H., James G., Gosset E. 
2011, MNRAS, 416, 817  
\bibitem[\protect\citeauthoryear{Sarajedini et al.}{2007}]{sarajedini07}
  Sarajedini A., Bedin L. R., Chaboyer B., Dotter  A., Siegel M.,
  Anderson J., Aparicio A., King I., Majewski S., Marin-Franch A.,
  Piotto G., Reid  I. N., Rosenberg A., Steven M. 2007, AJ, 133, 1658
\bibitem[\protect\citeauthoryear{Sesana, Haardt \& Madau}{2006}]{sesana06} Sesana A., 
Haardt F., Madau P. 2006, ApJ, 651, 392
\bibitem[\protect\citeauthoryear{Sesana}{2010}]{sesana10} Sesana A. 2010, ApJ, 719, 851  
\bibitem[\protect\citeauthoryear{Sesana et al.}{2012}]{sesana12} Sesana A., Sartore N., 
Devecchi B., Possenti A. 2012, MNRAS, 427, 502
\bibitem[\protect\citeauthoryear{Seth et al.}{2008}]{seth08} Seth A., Agueros M.,
Lee D., Basu-Zych A. 2008, ApJ, 678, 116
\bibitem[\protect\citeauthoryear{Shih et al.}{2010}]{shih10} Shih I. C., Kundu A., 
Maccarone T. J., Zepf S. E., Joseph T. D. 2010, ApJ, 721, 323
\bibitem[\protect\citeauthoryear{Sippel \& Hurley}{2013}]{sippel13} Sippel A. C., 
Hurley J. R. 2012, MNRAS, 430, 30
\bibitem[\protect\citeauthoryear{Sigurdsson \& Phinney}{1993}]{sigurdsson93a} Sigurdsson S.,
Phinney E. S. 1993, ApJ, 415, 631
\bibitem[\protect\citeauthoryear{Sigurdsson \& Hernquist}{1993}]{sigurdsson93b} Sigurdsson S.,
Hernquist L. 1993, Nature, 364, 423
\bibitem[\protect\citeauthoryear{Spergel et al.}{2003}]{spergel03} Spergel D. N. 2003,
ApJS, 148, 175
\bibitem[\protect\citeauthoryear{Spitzer}{1969}]{spitzer69} Spitzer L. Jr. 1969, ApJL, 158, L139
\bibitem[\protect\citeauthoryear{Spitzer}{1987}]{spitzer87} Spitzer L. Jr. 1987,
Dynamical Evolution of Globular Clusters (Princeton, NJ: Princeton Univ. Press)
\bibitem[\protect\citeauthoryear{Strader et al.}{2012a}]{strader12a} Strader J., Chomiuk L.,
Maccarone T. J., Miller-Jones J. C. A., Seth A. C., Heinke C. O., Sivakoff G. R. 2012, ApJ, 750, 27
\bibitem[\protect\citeauthoryear{Strader et al.}{2012b}]{strader12b} Strader J., Chomiuk L.,
Maccarone T. J., Miller-Jones J. C. A., Seth A. C. 2012, Nature, 490, 71
\bibitem[\protect\citeauthoryear{Umbreit \& Rasio}{2013}]{umbreit13} Umbreit S., Rasio F. A. 2013, 
ApJ, 768, 26
\bibitem[\protect\citeauthoryear{Valtonen \& Karttunen}{2006}]{valtonen06} Valtonen M., Karttunen H. 
2006, The Three-Body Problem (Cambridge: Cambridge University Press) 
\bibitem[\protect\citeauthoryear{van der Marel et al.}{2002}]{vandermarel02} van der Marel R. P., 
Gerssen J., Guhathakurta P., Peterson R. C., Gebhardt K. 2002, AJ, 124, 3255
\bibitem[\protect\citeauthoryear{van der Marel \& Anderson}{2010}]{vandermarel10} van der Marel R. P., 
Anderson J. 2010, ApJ, 710, 1063
\bibitem[\protect\citeauthoryear{Verbunt, Pooley \& Bassa}{2007}]{verbunt07}
  Verbunt F., Pooley D., Bassa C. 2007, in Dynamical Evolution of Dense Stellar
Systems, IAU Symp. 246, ed. E. Vesperini (Dordrecht: Reidel), 1
\bibitem[\protect\citeauthoryear{Vishniac}{1978}]{vishniac78} Vishniac E. T. 1978, ApJ, 223, 986
\bibitem[\protect\citeauthoryear{Wehner \& Harris}{2006}]{wehner06} Wehner E. H.,
Harris W. E. 2006, ApJL, 644, L17
\bibitem[\protect\citeauthoryear{Wrobel, Greene \& Ho}{2011}]{wrobel11} Wrobel J. M., 
Greene J. E., Ho L. C. 2011, AJ, 142, 113
\bibitem[\protect\citeauthoryear{Wyller}{1970}]{wyller70} Wyller A. A. 1970, ApJ, 160, 443
\bibitem[\protect\citeauthoryear{Zepf et al.}{2007}]{zepf07} Zepf S. E., Maccarone T. J., 
Bergond G., Kundu A., Rhode K. L., Salzer J. J. 2007, ApJL, 669, 69 
\end{thebibliography}
\end{document}